\documentclass[aps,reprint,twocolumns,pra,superscriptaddress,floatfix,notitlepage,nofootinbib]{revtex4-1}
\usepackage{amssymb,amsmath,amsfonts} 
\usepackage{mathtools}
\usepackage{physics}
\usepackage{array}
\usepackage{graphicx,xcolor}
\usepackage[colorlinks=true,citecolor=blue,linkcolor=red]{hyperref}
\usepackage{newtxtext}

\newcommand{\beq}{\begin{equation}}
\newcommand{\eeq}{\end{equation}}
\begin{document}

\title{Mechanism of dynamical phase transitions: The complex-time survival amplitude}

\author{\'{A}ngel L. Corps}
    \email[]{corps.angel.l@gmail.com}
    \affiliation{Instituto de Estructura de la Materia, IEM-CSIC, Serrano 123, E-28006 Madrid, Spain}
    \affiliation{Grupo Interdisciplinar de Sistemas Complejos (GISC),
Universidad Complutense de Madrid, Av. Complutense s/n, E-28040 Madrid, Spain}
    
\author{Pavel Str\'{a}nsk\'{y}}
    \email[]{stransky@ipnp.mff.cuni.cz}
    \affiliation{Institute of Particle and Nuclear Physics, Faculty of Mathematics and Physics,
Charles University, V Hole\v{s}ovi\v{c}k\'{a}ch 2, 18000 Prague, Czech Republic}

\author{Pavel Cejnar}
    \email[]{cejnar@ipnp.mff.cuni.cz}
    \affiliation{Institute of Particle and Nuclear Physics, Faculty of Mathematics and Physics,
Charles University, V Hole\v{s}ovi\v{c}k\'{a}ch 2, 18000 Prague, Czech Republic}

\date{\today} 

\begin{abstract}
Dynamical phase transitions are defined through non-analyticities of the survival probability of an out-of-equilibrium time-evolving state at certain critical times. They ensue from zeros of the corresponding survival amplitude. By extending the time variable onto the complex domain, we formulate the complex-time survival amplitude. The complex zeros of this quantity near the time axis correspond, in the infinite-size limit, to non-analytical points where the survival probability abruptly vanishes. Our results are numerically exemplified in the fully-connected transverse-field Ising model, which displays a symmetry-broken phase delimited by an excited-state quantum phase transition. A detailed study of the behavior of the complex-time survival amplitude when the characteristics of the out-of-equilibrium protocol changes is presented. The influence of the excited-state quantum phase transition is also put into context.
\end{abstract}

\maketitle

\section{Introduction}

Phase transitions represent qualitative changes in properties of physical systems induced by varying control parameters.
These changes become non-analytic in the infinite-size limit, which warrants their description as critical phenomena.
Historically, phase transitions were introduced in the framework of classical thermodynamics, but quantum physics created novel types of criticality driven by non-thermal parameters.
Within the framework of recent quantum technologies, phase transitions and critical phenomena play an important role in the discovery of new, exotic effects \cite{Baumann2010,Yuznashyan2006,Gring2012,Chu2020,Muniz2020}.

Phase transitions are commonly revealed by the response of a physical system to changes of some parameters.
In the limit of infinite system size, some special values of parameters defining so-called critical points may exhibit sudden changes in the response. 
In the case of systems coupled to a bath at constant temperature, a thermal phase transition (TPT) may occur at a critical temperature, when the partition function, and consequently many relevant thermodynamic quantities such as the free energy, do not behave smoothly.
Quantum systems at zero temperature may display signatures of a phase transition at some critical values of Hamiltonian control parameters, such as external fields or interaction strengths, giving rise to the zero-temperature quantum phase transition (QPT) that affect solely the ground-state properties of the system \cite{Sachdev1999}.
It was later discovered that phase transitions need not be restricted to the ground state; rather, signatures of criticality may be found also in highly excited parts of the energy spectrum. 
This phenomenon, originally observed in a class of quantum many-body systems with infinite-range interactions, was dubbed excited-state quantum phase transition (ESQPT) \cite{Cejnar2021,Cejnar2006,Caprio2008,Stransky2014,Stransky2015,Macek2019}.
The ESQPTs are defined as singularities of quantal spectra (in particular, non-analyticities of the level density and flow) on certain critical surfaces in the space composed of the Hamiltonian control parameters and the excitation energy. 
For a system with fixed control parameters, the existence of an ESQPT critical energy may cause a phase separation in the energy spectrum, the dynamical and thermodynamic properties of each phase differing significantly \cite{Corps2021,Puebla2013,Puebla2013b}. 
Over the years, our understanding of this generalization of the QPT concept has improved tremendously, and an active community has been formed around this phenomenon. 
For systems with a moderate number of degrees of freedom \cite{Stransky2016}, ESQPTs have a strong impact on the energy level dynamics \cite{Cejnar2008} and, consequently, entail a variety of relevant dynamical effects, including but not limited to anomalously enhanced decoherence \cite{Relano2008,PerezFernandez2009}, singular behavior in quench dynamics \cite{PerezFernandez2011,Santos2015,Lobez2016,Kloc2018,Stransky2021}, eigenstate localization \cite{Santos2016,PerezBernal2017}, anomalous thermalization \cite{Cejnar2017,Relano2018}, the induction of Schr\"{o}dinger-like cat states \cite{Corps2022PRA}, certain dynamical instabilities \cite{Bastidas2014} or irreversibility without energy dissipation \cite{Corps2022,Puebla2015}. 
ESQPTs are essentially a powerful quantum manifestation of the semiclassical limit of the system \cite{Wang2021}, since they are caused by critical points of the classical Hamiltonian flow \cite{Stransky2016}. 
Their consequences have been analyzed in a great variety of systems ranging from nuclear, molecular and atomic physics to quantum optics and, in general, condensed matter physics \cite{Ribeiro2007,Ribeiro2008,Perez2011b,Larese2013,Brandes2013,Relano2014,Bastarrachea2014,Puebla2016,Relano2016,Cejnar2016,Kloc2017,Khalouf2021,Khalouf2022,Gamito2022,Corps2022JPA}.
Extensions of ESQPTs to resonant states in the continuum \cite{Stransky2020,Stransky2021continuum} and to open systems weakly coupled to an environment \cite{RubioGarcia2022} have been proposed.

Although QPTs and ESQPTs entail dynamical consequences, they are really equilibrium, static phenomena, caused by some sort of singularity in the energy spectrum. Recently, the term dynamical phase transition (DPT) has been devised to refer to two new kinds of non-analytic effects. On the one hand, DPTs of the first type (DPT-I) \cite{Marino2022} denote some form of abrupt change in the long-time average of physically relevant observables, such as the total magnetization, after taking an initial state out of equilibrium by means of a quantum quench. These asymptotic values are somehow connected with pre-thermalization and define dynamical order parameters of the phase transition \cite{Eckstein2008,Moeckel2008,Eckstein2009,Sciolla2011,Zhang2017,Smale2019,Tian2020,Halimeh2017prethermalization,Sciolla2013}. On the other hand, DPTs of the second type (DPT-II) \cite{Heyl2013,Heyl2019} refer instead to non-analytic times in the return probability of the time evolved wavefunction to its initial state; this is therefore an inherently dynamical effect, taking place before the system has reached equilibration at all. There is no order parameter in the conventional sense linked to this kind of DPT, and the creation of different phases is not immediately apparent from the nature of the effect. In the seminal paper \cite{Heyl2013},  DPTs-II were originally proposed using the nearest-neighbor transverse-field Ising model, which may be solved exactly, by establishing a formal connection between the survival amplitude of an out-of-equilibrium state and the equilibrium, boundary partition function at complex inverse temperature $\beta=it$. A somewhat different quantity diagnosing DPTs-II in systems with $\mathbb{Z}_{2}$ symmetry-broken phases was also proposed \cite{Heyl2014}. Some systems are known to exhibit both kinds of DPTs \cite{Zunkovic2018,Lang2018,Puebla2020}, although the strict connection between them is still an open question \cite{Lang2018concurrence,Zunkovic2018,Lang2018,Weidinger2017,Hashizume2022,Sehrawat2021,Lerose2019,Zunkovic2015,Corps2022,Corps2022arxiv}. Inspired by the physics of fully-connected systems and the phases demarcated by an ESQPT, in \cite{Corps2022,Corps2022arxiv} DPTs-I and DPTs-II were argued to be triggered by the behavior of an operator which is constant only in some of the ESQPT phases. In this paper, we will be concerned with DPTs-II only, which we will henceforth simply denote DPTs. 

All kinds of phase transitions are only strictly realized in the infinite-size limit of the system under consideration. Nevertheless, precursors of such infinite-size abrupt changes can be found already for moderate system sizes. For the purposes of this work, we would like to recall the mechanism whereby phase transitions of real systems can be seen to arise from certain behaviors found by complexifying a control parameter, see, e.g., \cite{Cejnar2016}. For example, ground-state QPTs can be seen to arise from the crossing of the two lowest energy eigenvalues. These crossings are actually avoided at finite system-size, but with a vanishing gap as the size increases. QPTs may be viewed through the prism of eigenvalue degeneracies in a complex-extended plane of a control parameter $\Lambda$. After the complexification $\Lambda\in\mathbb{R}\to(\textrm{Re}\,\Lambda+i\,\textrm{Im}\,\Lambda)\equiv\Lambda\in\mathbb{C}$, the complex eigenvalues of the now non-Hermitian Hamiltonian may exhibit so-called exceptional points \cite{Heiss2012}, i.e., complex degeneracies, at certain values of $\Lambda$. Naturally, this complex-valued control parameter is not always physically relevant; however, if an exceptional point occurs near the real axis, $\textrm{Im}\,\Lambda\approx0$, an avoided crossing is induced in the real spectrum at $\textrm{Re}\,\Lambda$~\cite{Cejnar2005,Cejnar2007,Stransky2018}. Similarly, a thermal phase transition can be seen to ensue from the zeros of the complex-extended partition function, where it is the temperature $\beta$ which becomes complex. This is the Yang-Lee approach to thermal phase transitions \cite{Yang1952,Lee1952,Fisher1965,Grossmann1969,Itzykson1983,Derrida1991,Borrmann2000,Wei2014}. Based on these ideas, in this work we propose a related generalization of the survival amplitude, whose zeros in the complex plane define DPTs. We analyze the usefulness of this approach, related to the previously explained conception of the boundary partition function as a survival amplitude at complex temperature \cite{Heyl2013}. Use of Loschmidt cumulants has also been made in the determination of DPTs \cite{Peotta2021, Brange2022}. 

This paper is organized as follows. In Sec.~\ref{sec:complex-time} we present the complex-time survival amplitude, which is the main mathematical object of our work. The model that we employ to exemplify our results and the quench procedure used is introduced in Sec.~\ref{sec:model}. The main results follow in Sec.~\ref{sec:results}, where we focus on the identification of zeros of the complex-time survival amplitude, we analyze the scaling of number of zeros with system size as well as the dependence on the initial condition and energy of the quenched state. The influence of ESQPTs is considered at the end of this section. We finally conclude in Sec.~\ref{sec:conclusions}.

\section{Complex-time survival amplitude}\label{sec:complex-time}

Let us consider a bound quantum system with Hamiltonian $\hat{H}(\Lambda^\mu)$ depending on a set $\{\Lambda^\mu\}_{\mu=0,1,2\dots}$ of some control parameters.
We will describe the dynamics of this system after a quantum quench, i.e., after a sudden parameter change $\Lambda^\mu_{i}\to\Lambda^\mu_{f}$.
The initial state $\ket{\Psi_{0}(\Lambda^\mu_{i})}$ at time ${t=0}$ is chosen as a stationary state of the initial Hamiltonian $\hat{H}(\Lambda^\mu_{i})$, that is, as one of its discrete eigenstates or as a superposition of several degenerate eigenstates. 
After the quench, the system is no more stationary.
Its wavefunction evolves according to the time-independent Sch\"{o}dinger equation, at time $t\geq 0$ being expressed as
\beq
\begin{split}\label{eq:schrodinger}
\bigl|\Psi_{t}(\Lambda^\mu_{f})\bigr\rangle &=e^{-i\hat{H}(\Lambda^\mu_{f})t}\ket{\Psi_{0}(\Lambda^\mu_{i})}\\&=\sum_{n}c_{n}e^{-iE_{n}(\Lambda^\mu_{f})t}\bigl|E_{n}(\Lambda^\mu_{f})\bigr\rangle,
\end{split}
\eeq
where $E_{n}(\Lambda^\mu_{f})$ and $|E_{n}(\Lambda^\mu_{f})\rangle$, respectively, represent discrete eigenvalues and eigenvectors of the final Hamiltonian $\hat{H}(\Lambda^\mu_{f})$ and $c_{n}\equiv \langle E_{n}(\Lambda^\mu_{f})|\Psi_{0}(\Lambda^\mu_{i})\rangle$ are normalized expansion coefficients of the initial state in the final Hamiltonian  eigenbasis.

The survival amplitude of the initial state in the evolving state~\eqref{eq:schrodinger} reads 
\beq\label{eq:survivalamplitude}
A(t)=\bigl\langle\Psi_{0}(\Lambda^\mu_{i})\bigr|\Psi_{t}(\Lambda^\mu_{f})\bigr\rangle=\sum_{n}|c_{n}|^{2}e^{-it E_{n}(\Lambda^\mu_{f})}.
\eeq
It is completely determined by the final Hamiltonian eigenvalues $E_{n}(\Lambda^\mu_{f})$ and the corresponding population probabilities $|c_n|^2$. 
The \textit{survival probability\/} of the initial state after time $t$ is trivially given by $\mathcal{P}(t)=|A(t)|^{2}$. 
The notion of dynamical phase transitions was initiated in its DPT-II incarnation, namely as zeros of the survival probability and amplitude at some time instants ${t=t_*}$~\cite{Heyl2013,Heyl2014}.
The condition $\mathcal{P}(t_{*})=0=A(t_{*})$ corresponds to exact orthogonality (zero overlap) of the evolving wavefunction to its initial form and can be seen as a momentary loss of the system's memory.
This kind of dynamical criticality is in our focus below. 

The reason why the above zeros of the survival amplitude are considered as a kind of quantum critical effect results from the striking similarity to zeros of the partition functions, which constitute a well established description of the TPTs.
Recall that thermal properties of general systems in canonical thermodynamics are derived from the partition function $Z(\beta)=\sum_{n}g_{n}e^{-\beta E_{n}}$, where ${\beta=1/T}$ is the inverse temperature and $g_{n}\in\mathbb{N}$ is a degeneracy factor associated with the $n$th energy level.
By definition, $Z(\beta)$ in any finite system cannot be zero at any real $\beta$, but it can take values very close to zero, which then implies singular behavior of thermodynamic observables.  
The TPTs can be understood through true zeros of a complexified partition function
\beq\label{eq:Zthermal}
Z(z)=\sum_{n}g_{n}e^{-z E_{n}},
\eeq
where the inverse temperature is extended to the complex plane: $z=\Re(z)+i\Im(z)\in\mathbb{C}$.
If ${Z(z_0)=0}$ at a point $z_0\in\mathbb{C}$ whose imaginary part $\Im(z_0)$ drops to zero with increasing size of the system, the real partition function $Z(\beta)$ in the infinite-size limit generates a TPT at ${\beta=\Re(z_0)}$.
The advantage of this approach, which was introduced in Ref.\,\cite{Yang1952,Lee1952} and further elaborated, e.g., in Refs.\,\cite{Fisher1965,Grossmann1969,Itzykson1983,Derrida1991,Borrmann2000,Wei2014}, is the possibility to trace precursors of the critical behavior in strictly finite systems.

The survival amplitude~\eqref{eq:survivalamplitude} formally resembles the partition function~\eqref{eq:Zthermal} at ${z=-it}$ and can be treated in a completely analogous way.
Hiding the dependence of energies on $\Lambda^\mu$, we define 
\beq\label{eq:Z}
\mathcal{Z}(z)=\sum_{n}|c_{n}|^{2}e^{-z E_{n}}=\sum_{n}|c_{n}|^{2}e^{-\beta E_{n}}e^{-itE_{n}},
\eeq
where $z=\beta + it\in\mathbb{C}$.
There are two ways how to look at this quantity.
First, it can be interpreted as a partition function of a fictitious system with ${g_{n}\propto|c_{n}|^{2}}$ in complexified inverse temperature. 
More precisely, if $d$ denotes the total number of states (we assume that our system has a finite spectrum), then $\mathcal{Z}(z)$ is a partition function divided by $d$ of a system with non-integer degeneracy factors ${g_n=|c_{n}|^{2}}d$ satisfying ${\sum_n g_n=d}$.
Zeros of $\mathcal{Z}(z)$ near the $\beta$ axis (real axis of $z$) would imply precursors of a TPT in that system.

The second interpretation of function~\eqref{eq:Z} is in terms of the survival amplitude $A(t)$ from Eq.~\eqref{eq:survivalamplitude} in a complexified time.
Zeros of this function near the $t$ axis (imaginary axis of $z$) generate precursors of critical behavior in the time domain.
In particular, if ${\mathcal{Z}(z_0)=0}$ at a point $z_0\in\mathbb{C}$ whose real part $\Re(z_0)$ drops to zero with increasing size of the system, the real survival amplitude $A(t)$ in the infinite-size limit generates a DPT at time ${t=\Im(z_0)\equiv t_0}$.

The quantity~\eqref{eq:Z} with any value of $\beta$ is proportional to the survival amplitude of a system with modified population coefficients. 
Indeed, defining normalized population probabilities $\widetilde{p}_{n}(\beta)\equiv |c_{n}|^{2}e^{-\beta E_{n}}/\sum_{k}|c_{k}|^{2}e^{-\beta E_{k}}$, we see that the survival amplitude of the corresponding state is given by
\beq\label{eq:A}
\mathcal{A}(\beta,t)=\sum_{n}\widetilde{p}_{n}(\beta)e^{-itE_{n}}=\frac{\mathcal{Z}(\beta+it)}{\mathcal{Z}(\beta+i0)}.
\eeq
This formula for ${\beta=0}$ represents the original survival amplitude~\eqref{eq:survivalamplitude}, while the cases with ${\beta\neq 0}$ yield survival amplitudes of quantum states with enhanced populations of low-energy (${\beta>0}$) or high-energy (${\beta<0}$) parts of the spectrum.
Therefore, the zero of $\mathcal{Z}(z)$ at any ${z_0=\beta_0+it_0}$ represents an actual time singularity at ${t=t_0}$ of a system with occupation probabilities $\widetilde{p}_{n}(\beta_0)$.

Since zeros of $\mathcal{Z}(z)$ are determined by two independent conditions on the vanishing real and imaginary parts, they form isolated points in the complex plane of variable $z$.
The number of these points is in general infinite. 
The function~\eqref{eq:Z} is differentiable in both real variables $\beta$ and $t$, and allows for regular Taylor expansion in a vicinity of points $\mathcal{Z}(z_0)=0$.
A~direct way to locate these points makes use of the nodal lines of $\mathcal{Z}(z)$, that is, the contour lines obtained by imposing separately the conditions $\Re\mathcal{Z}(z)=0$ and $\Im\mathcal{Z}(z)=0$. 
Zeros of $\mathcal{Z}(z)$ appear at intersections of these lines.
Numerous examples will be shown below.\\

\section{Model}\label{sec:model}

\subsection{Hamiltonian}

Although the main ideas of this paper are general, for our numerical calculations we will choose the paradigmatic fully-connected transverse-field Ising model~\cite{Sachdev1999}. The system is described by the Hamiltonian
\beq\label{eq:isingchain}
\hat{H}=-\frac{\lambda}{4N}\sum_{i,j=1}^{N}\hat{\sigma}_{i}^{x}\hat{\sigma}_{j}^{x}+\frac{h}{2}\sum_{i=1}^{N}\hat{\sigma}_{i}^{z},
\eeq
with control parameters $\{\Lambda^{\mu}\}=\{h,\lambda\}$, where $h$ is a transverse magnetic field and $\lambda$ is the interaction strength, $N$ is the number of sites, and $\hat{\sigma}_{i}^{x,y,z}$ are the Pauli matrices of the spin-$1/2$ particle at site $i$. Instead of first-neighbor or otherwise finite-range interactions, we allow for collective (infinite-range) interactions, whereby each spin in the chain interacts with all other spins. As a consequence, $\hat{H}$ describes a fully-connected system. Thus, the Jordan-Schwinger representation for the collective spin operators $\hat{J}_{x,y,z}=\frac{1}{2}\sum_{i=1}^{N}\hat{\sigma}_{i}^{x,y,z}$ may be used to cast Eq.~\eqref{eq:isingchain} in the simpler form
\beq\label{eq:lipkin}
\hat{H}=-\frac{\lambda}{N}\hat{J}_{x}^{2}+h\hat{J}_{z}.
\eeq
This Hamiltonian is a special case of the famous Lipkin-Meshkov-Glick (LMG) model \cite{Lipkin1965}, originally formulated in the fermionic language as a schematic example of the nuclear shell model. Far from remaining within the field of nuclear physics, the LMG model has been revealed as a powerful testbed for a range of different physical phenomena, including quantum phase transitions \cite{Sehrawat2021,Castanos2006,Relano2008,Homrighausen2017,Stransky2018,Lang2018concurrence,Cejnar2016}, and it has been experimentally realized with cold atoms \cite{Muniz2020}. 

It is clear from Eq.~\eqref{eq:lipkin} that the total spin operator $\hat{\mathbf{J}}^{2}$ is a conserved quantity, $[\hat{H},\hat{\mathbf{J}}^{2}]=0$. Therefore, we may separate the full Hamiltonian matrix according to its eigenvalues, $j(j+1)$, $j=0,1,...,N/2$, in $j$-symmetry sectors, with dimension $2j+1$ each, implying a huge Hilbert space dimension reduction, from exponential to linear. Individual $j$-sectors have generally a large number of replicas differing by qubit permutation symmetry, but we choose the unique fully symmetric sector with the maximal $j=N/2$, which includes the ground-state of the system. The operator $\hat{\Pi}\equiv e^{i\pi (j+\hat{J}_{z})}$, called parity, is yet another symmetry of Eq.~\eqref{eq:lipkin}, with a discrete spectrum, $\textrm{Spec}\,(\hat{\Pi})=\{-1,+1\}$, providing discrete quantum numbers according to $\hat{\Pi}\ket{E_{n,\pm}}=\pm \ket{E_{n,\pm}}$, where $\{\ket{E_{n,\pm}}\}_{n}$ are the positive/negative parity eigenvectors of Eq.~\eqref{eq:lipkin}. Precisely due to the collective nature of the LMG Hamiltonian \cite{Cejnar2021}, the infinite-size limit,  $N\to\infty$, can be shown to correspond exactly with the semiclassical limit, $j\to\infty$, which does \textit{not} hold true in case that Eq.~\eqref{eq:isingchain} is restricted to any finite-range interaction.

The mean-field solution may be obtained, e.g., by use of the Bloch coherent states
\beq
\ket{\omega}=\left(\frac{1}{1+|\omega|^{2}}\right)^{j}e^{\omega \hat{J}_{+}}\ket{j,-j},
\eeq
where $\ket{j,m}$ is the $\hat{J}_{z}$ eigenbasis, $\hat{J}_{z}\ket{j,m}=m\ket{j,m}$, and $\omega=(Q+iP)/\sqrt{4-P^{2}-Q^{2}}$, with $Q$ and $P$ real, are canonical variables constrained as $0\leq Q^{2}+P^{2}\leq 4$, implying compactness of the classical phase space. A semiclassical representation of the quantum Hamiltonian Eq.~\eqref{eq:lipkin} is therefore given by the following intensive ($j$-independent) function:
\begin{align}\label{eq:classical}
H(Q,P)&\equiv \frac{\bra{\omega}\hat{H}\ket{\omega}}{j}\\&=-h+\frac{h}{2}(Q^{2}+P^{2})-\frac{\lambda}{8}Q^{2}(4-P^{2}-Q^{2}).\nonumber
\end{align}
The classical system described by $H(P,Q)$ has a single degree of freedom, and therefore it is integrable. Its energy scale is given by the intensive quantity $\epsilon=E/j$, where $E$ denotes the actual, $j$-dependent eigenvalues of the quantum model. Here, we briefly summarize the main semiclassical features for completeness.

\begin{figure}[h!]
    \centering
    \begin{tabular}{c}
\hspace{-0.5cm}\includegraphics[width=0.45\textwidth]{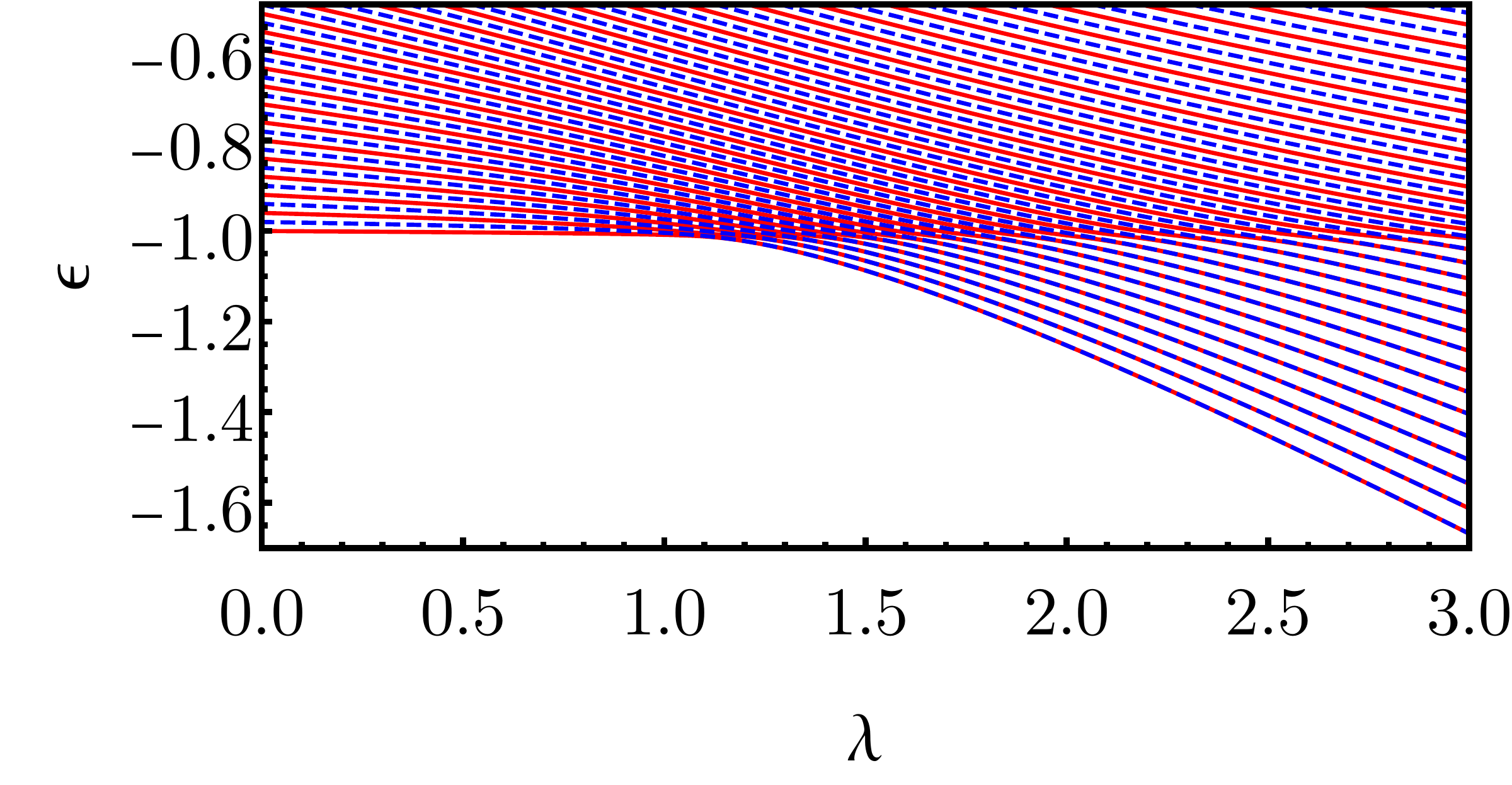} \\

\begin{tabular}{cc}
\hspace{-0.4cm}\includegraphics[width=0.25\textwidth]{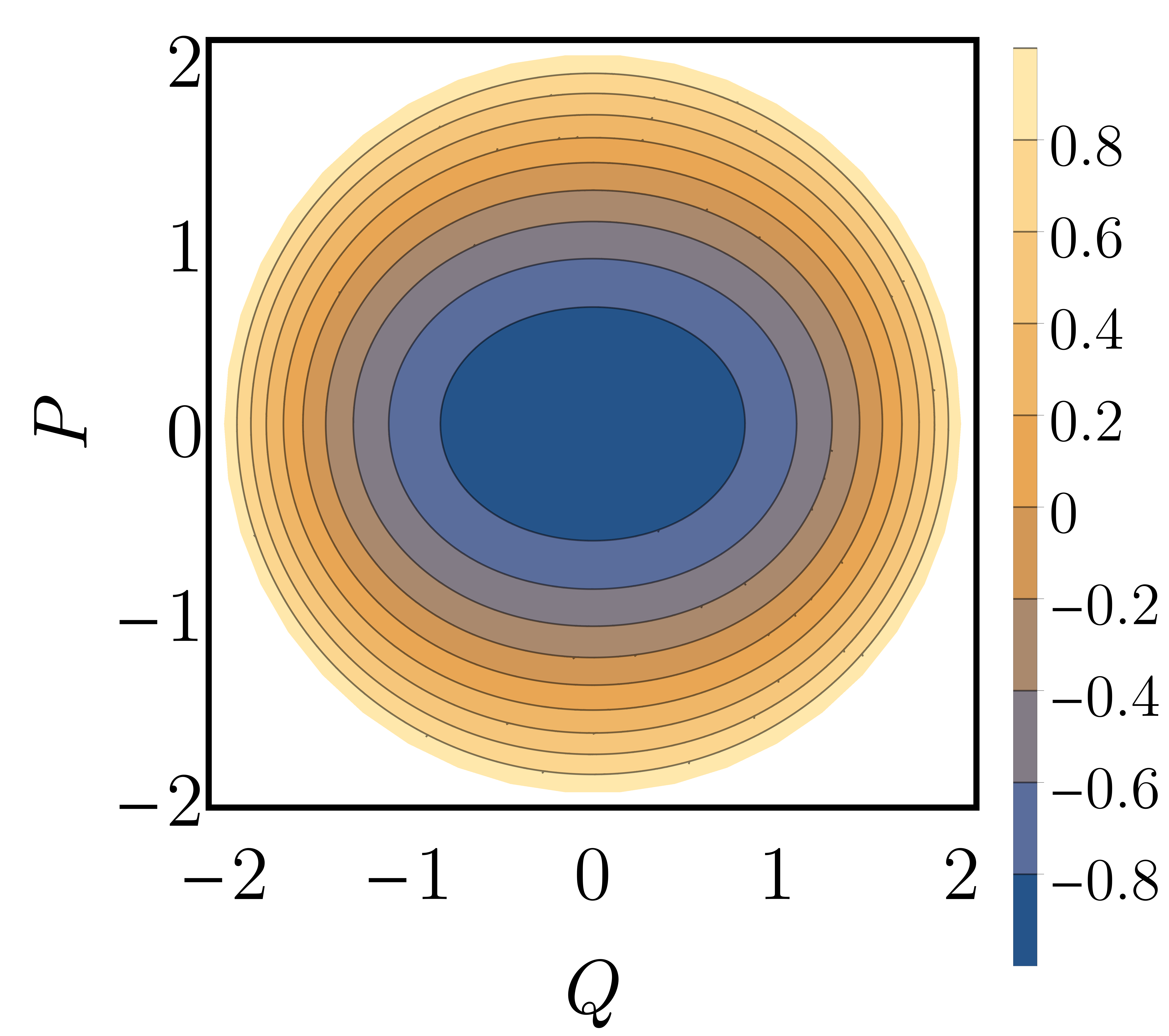} & 
\hspace{-0.2cm}\includegraphics[width=0.25\textwidth]{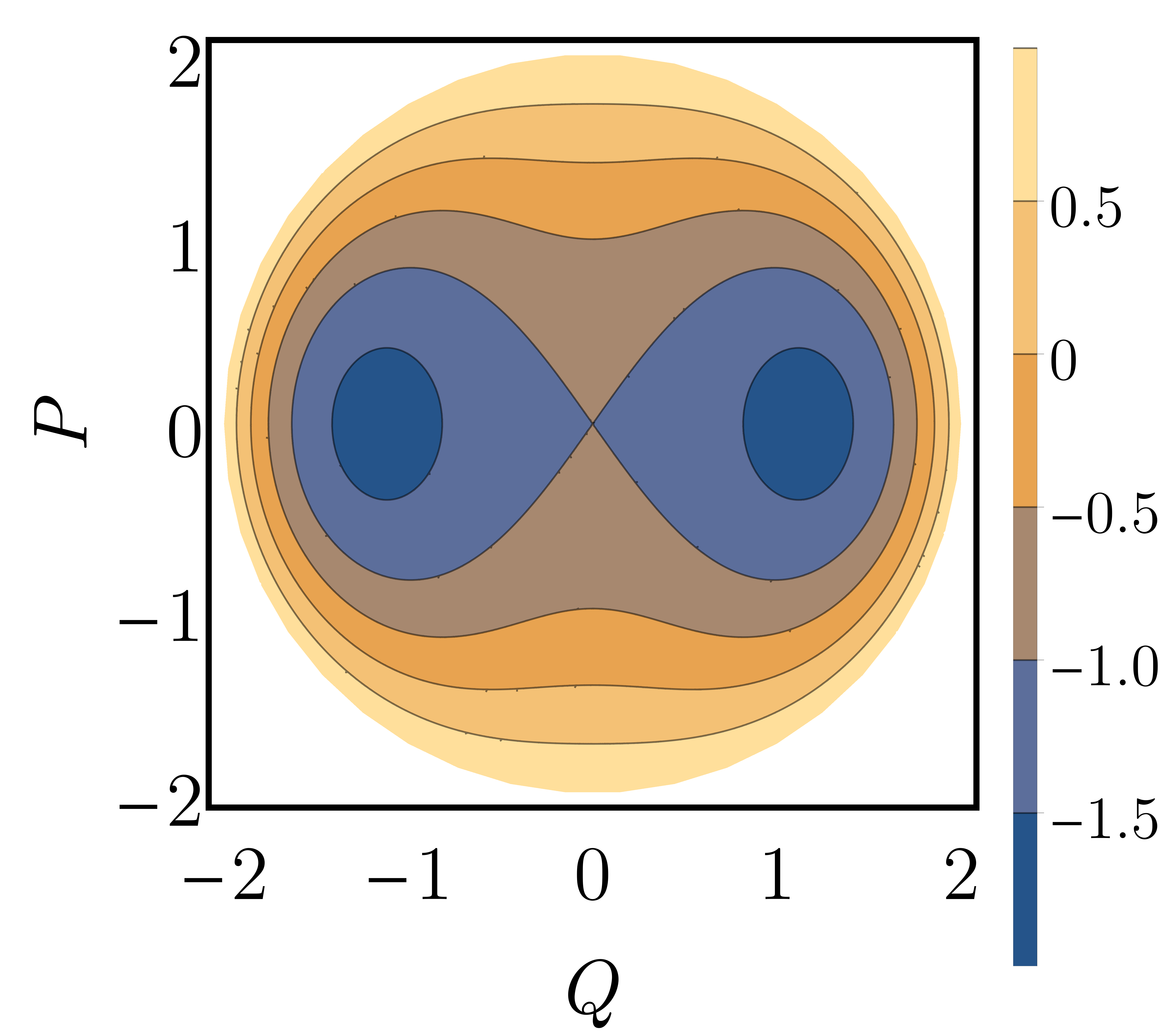}\\
\end{tabular}\\

\hspace{-0.5cm}\includegraphics[width=0.45\textwidth]{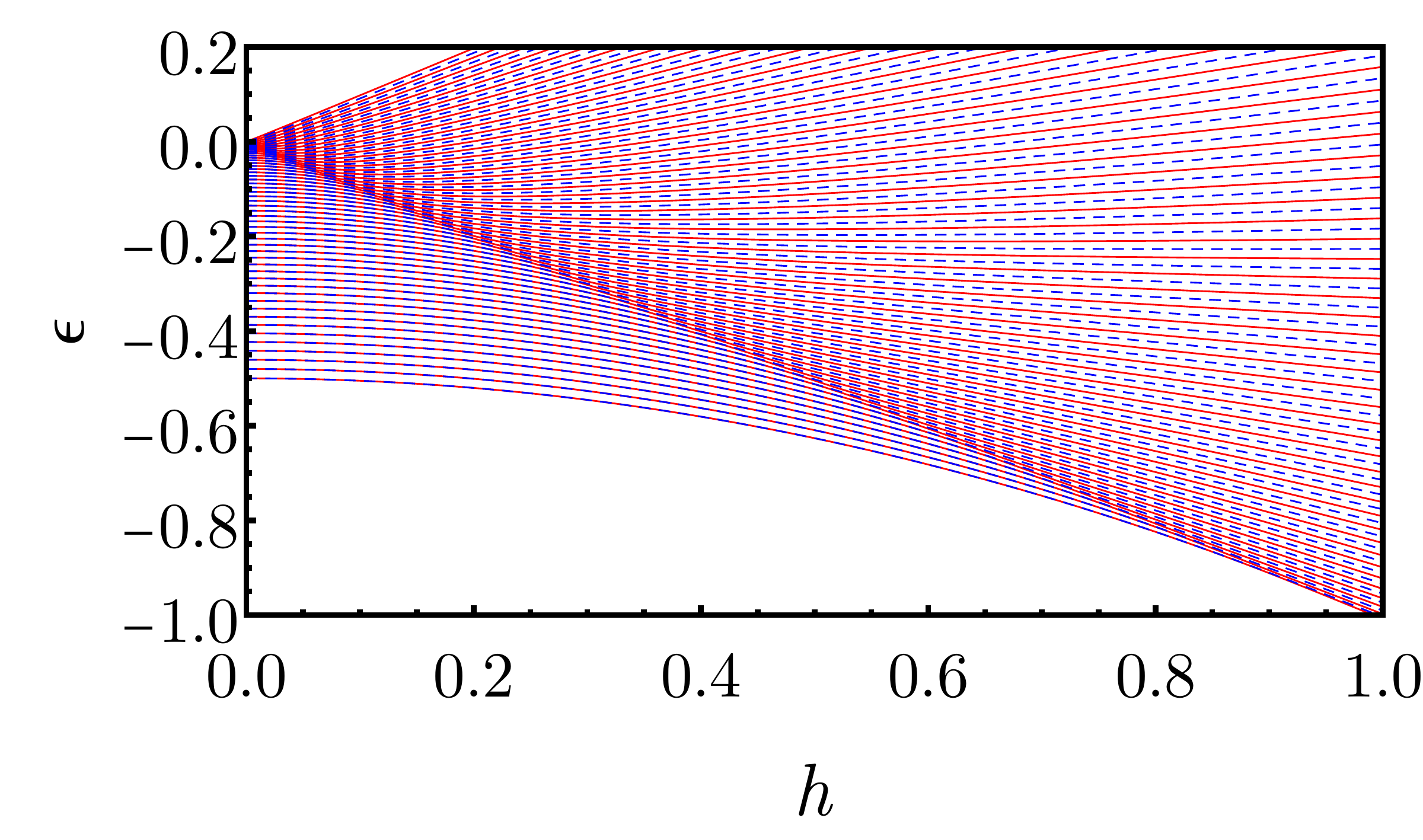}
    \end{tabular}
    \caption{Level flow diagram (top) of the infinite-range transverse field Ising model Eq.~\eqref{eq:lipkin}. Red solid lines represent eigenvalues of positive parity, $E_{n,+}$, while blue dashed lines represent eigenvalues of negative parity, $E_{n,-}$. System size is $j=50$. Classical phase space for $\lambda=1/2<\lambda_{c}$ (middle, left) and $\lambda=3>\lambda_{c}$ (middle, right), obtained from the contour lines of the classical Hamiltonian Eq.~\eqref{eq:classical}. In all cases, $h=1$. Level flow diagram (bottom) for $\lambda=1$ and $h\in[0,1]$, with the same color code as before. }
    \label{fig:lipkinphase}
\end{figure}

The model exhibits a second-order ground-state quantum phase transition (QPT) at the critical coupling strength $\lambda_{c}=h$, inducing a structural change from single-well to double-well in the classical potential, as observed in Fig.~\ref{fig:lipkinphase}. The energy minimum in the infinite-size limit, calculated from~\eqref{eq:classical}, is
\begin{equation}
    \label{eq:minimum}
    \epsilon_{\mathrm{min}}=\begin{cases}
        -h & \lambda<\lambda_{c}=h,\\
        -\frac{1}{2}\left(\lambda + \frac{h^{2}}{\lambda}\right) & \lambda>\lambda_{c}=h.
    \end{cases}
\end{equation}
For $\lambda>\lambda_{c}$, the system also undergoes an ESQPT~\cite{Cejnar2021,Caprio2008,Stransky2014} at the critical energy $\epsilon_{c}=-h$ \cite{Ribeiro2007,Ribeiro2008,PerezFernandez2009}, where, in systems with a single degree of freedom, the tight avoided crossing of eigenvalues translates into a logarithmic divergence of the density of states. For $\epsilon<\epsilon_{c}$, eigenlevels of different parity are degenerate in the infinite-size limit, $\epsilon_{n,+}=\epsilon_{n,-}$, while such degeneracy is lifted for $\epsilon>\epsilon_{c}$; in other words, eigenstates below the ESQPT are symmetry-broken. At the classical level, such degeneracies can be seen to arise from the double-well potential, which allows for symmetric, disconnected wells below $\epsilon_{c}$. It is clearly observed in the classical portrait of Fig.~\ref{fig:lipkinphase} (middle, right) that all trajectories $(Q(t),P(t))$ with energy $H(Q(t),P(t))<\epsilon_{c}$ conserve the sign of the canonical position, $\textrm{sign}\,Q(t)=\pm 1$, since they are bound to either the left or right well. Trajectories with energy $H(Q(t),P(t))>\epsilon_{c}$ do not satisfy this condition, and thus $\textrm{sign}\,Q(t)$ is no longer conserved. 

It was proposed in~\cite{Corps2021} that the previous classical ideas may be carried over to the quantum realm by establishing a quantum-classical correspondence of constant classical dynamical functions and quantum operators. Since classically $j_{x}(Q,P)=\bra{\omega}\hat{J}_{x}\ket{\omega}/j=Q\sqrt{4-P^{2}-Q^{2}}/2\propto Q$, the sign of $Q(t)$ and the sign of $j_x (t)$ coincide. Thus, the quantum operator
\beq\label{eq:C}
\hat{\mathcal{C}}=\textrm{sign}(\hat{J}_{x})
\eeq
defines a discrete $\mathbb{Z}_{2}$ symmetry with eigenvalues $\textrm{Spec}\,(\hat{\mathcal{C}})=\{\pm 1\}$. This operator is constant only below $\epsilon_c$, so it is an instance of so-called partial symmetries~\cite{Leviatan2011}. Given an arbitrary quantum state $\ket{\varphi(t)}$, the expectation value $\bra{\varphi(t)}\hat{\mathcal{C}}\ket{\varphi(t)}\in[-1,1]$ indicates the region of the classical phase space occupied by such quantum state: completely within the right ($+1$) or left ($-1$) wells, or a superposition of both (between $-1$ and $+1$). For details on this operator, see~\cite{Corps2021,Corps2022PRA,Corps2022JPA,Corps2022}.

\begin{figure*}[t!]
    \centering
    \begin{tabular}{c c c}
\includegraphics[width=0.33\textwidth]{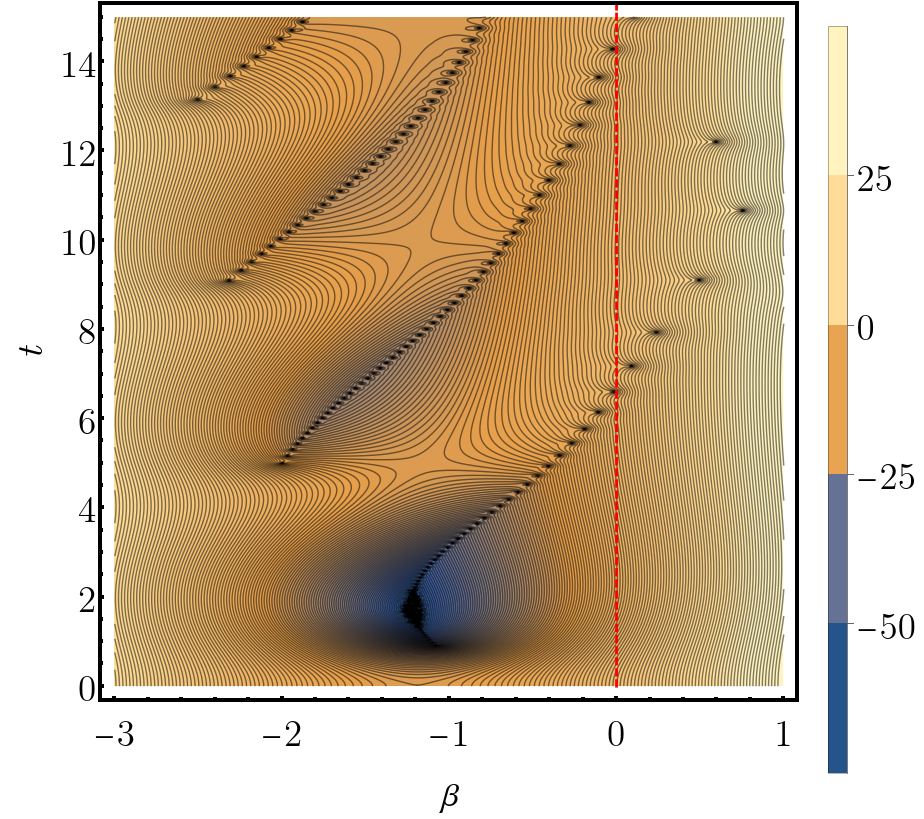} & \includegraphics[width=0.33\textwidth]{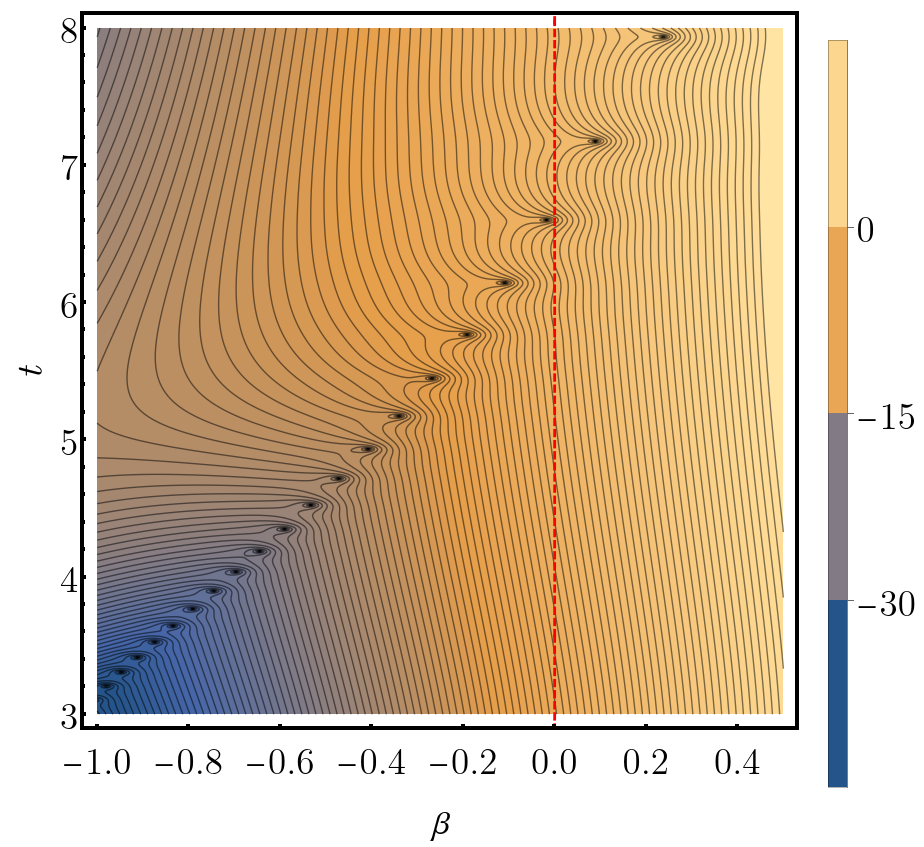} & \includegraphics[width=0.29\textwidth]{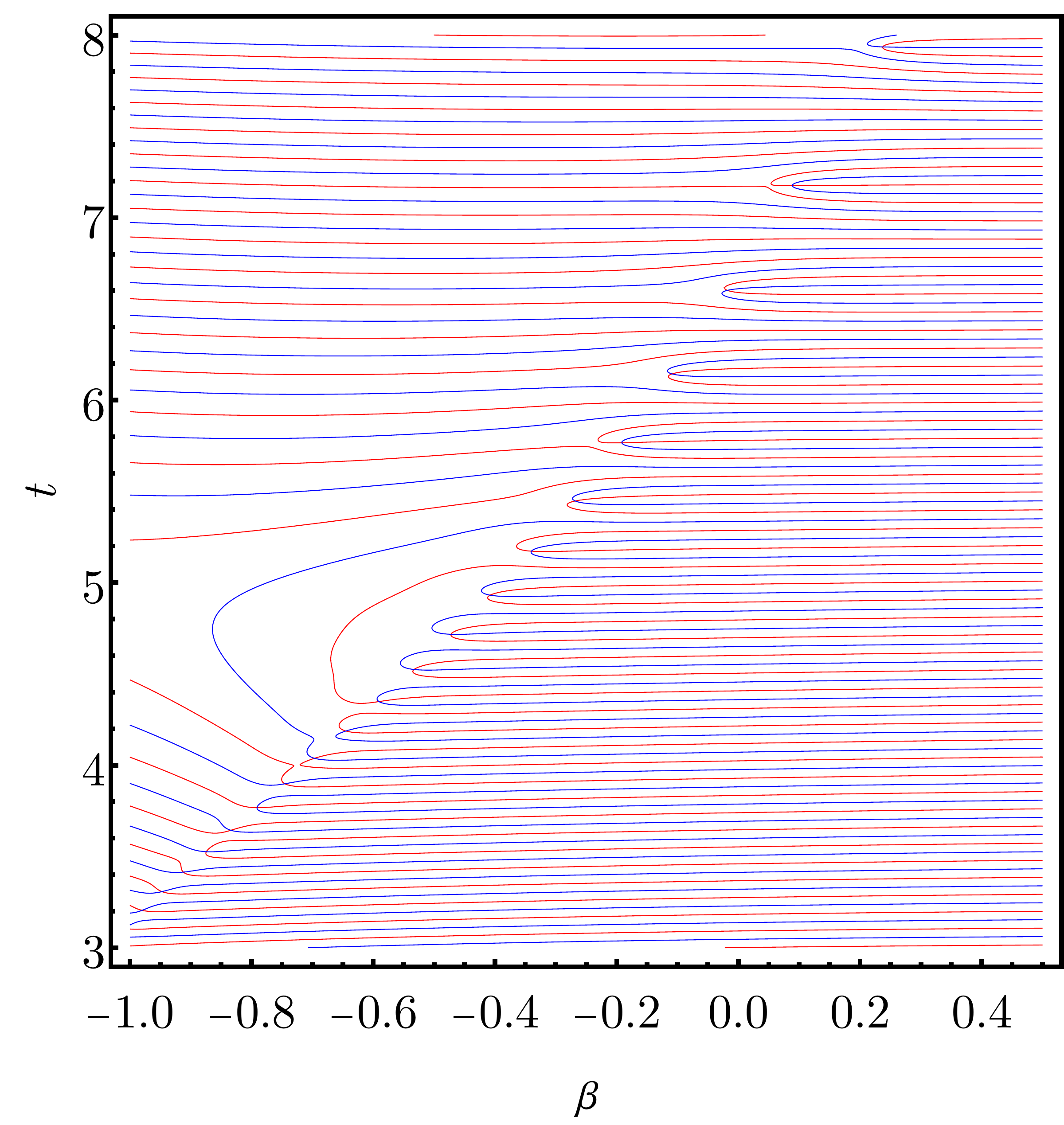}   
    \end{tabular}
    \caption{Contour plot of $\log_{10}|\mathcal{Z}(\beta+it)|^{2}$ in the $(\beta,t)$ plane (left). The red, vertical, dashed line signals the $t$-axis ($\beta=0$). Contour plot of $\log_{10}|\mathcal{Z}(\beta+it)|^{2}$ in a smaller window (middle) together with the corresponding nodal lines (right); red and blue lines correspond to the real and imaginary part of $\mathcal{Z}(\beta,t)$, respectively. A quench $\lambda_{i}=7\to\lambda_{f}=1.5$ has been performed, and the magnetic field is fixed at $h=1$. System size is $j=30$, and the parameters in the initial state~\eqref{eq:initialstate} are $\alpha=1/2$ and $\phi=0$.}
    \label{fig:Zpresentation}
\end{figure*}

\subsection{Quenches}
\label{sec:qu}

To prepare non-stationary states of the LMG model, we use the quantum quench method.
As explained above, a `quench' is a sudden jump of Hamiltonian control parameters performed in a system that was previously prepared in a stationary state of the Hamiltonian at the initial parameter value.
The jump generates a dispersed population of excited eigenstates of the final Hamiltonian, which induces the subsequent nontrivial evolution.
Such a non-thermal excitation can be---at least to a certain extent---purposely localized in selected excitation domains and may therefore be used to probe dynamical response of the system in these domains.

In the present model, the initial state $\ket{\Psi_{0}(\lambda_{i})}$ (if not stated otherwise, the other Hamiltonian parameter is fixed at the value $h=1$ and not explicitly mentioned in the formulas) is chosen to be a superposition of the nearly degenerate eigenstates of different parity that form the symmetry-broken ground state of Hamiltonian~\eqref{eq:lipkin} at a certain initial coupling strength $\lambda_{i}>\lambda_{c}$. 
In particular, we have
\beq\label{eq:initialstate}
\ket{\Psi_{0}(\lambda_{i})}=\sqrt{\alpha}\ket{E_{0,+}(\lambda_{i})}+e^{i\phi}\sqrt{1-\alpha}\ket{E_{0,-}(\lambda_{i})}.
\eeq
In this formula, $\alpha\in[0,1]$ determines population probabilities for both parity eigenstates and $\phi\in[0,2\pi)$ represents a relative phase.
In the classical phase space, the location of such an initial state can be derived exactly. The $\hat{\mathcal{C}}$ operator only connects eigenstates of opposite parity \cite{Corps2021}, its expectation value being $\bra{E_{n,\pm}}\hat{\mathcal{C}}\ket{E_{n,\mp}}=1$ in the spectral region where it is conserved and $\bra{E_{n,\pm}}\hat{\mathcal{C}}\ket{E_{n,\pm}}=0$ for states of the same parity. Thus, for the initial state Eq. \eqref{eq:initialstate}, this immediately brings
\beq\label{eq:classic}
\bra{\Psi_{0}(\lambda_{i})}\hat{\mathcal{C}}\ket{\Psi_{0}(\lambda_{i})}=2\sqrt{\alpha(1-\alpha)}\cos\phi.
\eeq
Strictly speaking, the state~\eqref{eq:initialstate} is stationary with respect to the initial Hamiltonian only in the infinite-size limit, when the degeneracy of the parity doublet becomes exact.

The quenches in the parameter space of Hamiltonian~\eqref{eq:lipkin} will be performed mostly in $\lambda$, but at last also in $h$.
Returning to the general Hamiltonian expression $\hat{H}(\Lambda^{\mu})$ with a set of real parameters $\Lambda^{\mu}$ ($\mu=0,1,2,\dots$), we can make an identification $(h,\lambda)\equiv(\Lambda^0,\Lambda^1)$. 
Our LMG Hamiltonian is obviously linear in both these parameters.
To elucidate some common properties of quenches in such linear systems, let us consider a general Hamiltonian
\beq\label{eq:Hgen}
{\hat{H}(\Lambda^{\mu})=\Lambda^0\hat{G}_0+\Lambda^1\hat{G}_1+\dots\equiv\Lambda^\mu\hat{G}_{\mu}}
\eeq
with parameters $\Lambda^{\mu}$ weighting some generally non-commuting Hamiltonian components~$\hat{G}_{\mu}$.
Any jump in the parameter space of such a~system can be expressed as ${\hat{H}_f=\hat{H}_i+\Delta^\mu\hat{G}_\mu}$, where ${\hat{H}_i\equiv\hat{H}(\Lambda_i^{\mu})}$ and ${\hat{H}_f\equiv\hat{H}(\Lambda_f^{\mu})}$ are the initial and final Hamiltonians, respectively, and $\Delta^\mu={\Lambda^\mu_f-\Lambda^\mu_i}$ is the parameter change (summation convention is used for Greek indices).
We denote an average value of a general operator $\hat{A}$ in the initial state as $\bra{\Psi_0(\Lambda^\mu_i)}\hat{A}\ket{\Psi_0(\Lambda^\mu_i)}\equiv\langle\hat{A}\rangle_i$ and the corresponding variance as ${\langle\hat{A}^2\rangle_i-\langle\hat{A}\rangle_i^2}\equiv\langle\!\langle\hat{A}^2\rangle\!\rangle_i$.
The energy average of the system after the quench is then given by 
\beq\label{eq:eneav}
\langle\hat{H}_f\rangle_i=\langle\hat{H}_i\rangle_i+\Delta^\mu\langle\hat{G}_\mu\rangle_i,
\eeq
while the energy variance reads
\begin{eqnarray}
\label{eq:enevar}
\langle\!\langle\hat{H}_f^2\rangle\!\rangle_i=&&
\langle\!\langle\hat{H}_i^2\rangle\!\rangle_i
\\
&+&2\Delta^\mu\bigl(\Re\langle\hat{H}_i\hat{G}_\mu\rangle_i-\langle\hat{H}_i\rangle_i\langle\hat{G}_\mu\rangle_i\bigr)
\nonumber\\
&+&\Delta^\mu\Delta^\nu\bigl(\langle\hat{G}_\mu\hat{G}_\nu\rangle_i-\langle\hat{G}_\mu\rangle_i\langle\hat{G}_\nu\rangle_i\bigr).
\nonumber
\end{eqnarray}
Formula~\eqref{eq:eneav} determines the center of mass of the initial-state energy distribution in the final Hamiltonian eigenbasis.
This distribution, determined by energies $E_n(\Lambda^{\mu}_f)$ and the corresponding population probabilities $|c_n|^2$, is called the local density of states (LDOS).
The knowledge of the final energy average is important for the design of quenches that excite the system to the desired energy domain (e.g., to a vicinity of an ESQPT).
Formula \eqref{eq:enevar} measures the squared width of the energy distribution.
This sets an `energy resolution' of the quench-based excitation procedure and determines the characteristic time of the initial decay of the survival probability, which is inversely proportional to the width.  

In the case of our LMG Hamiltonian, the energy average~\eqref{eq:eneav} after an arbitrary quench $(h_{i},\lambda_{i})\rightarrow(h_{f},\lambda_{f})$ from the initial state~\eqref{eq:initialstate} can be calculated exactly in the ${j\to\infty}$ limit.
From the Helmann-Feynman formula, $\langle\hat{G}_{\mu}\rangle_{i}=\partial E_{0}/\partial\Lambda^{\mu}|_{\Lambda^{\mu}=\Lambda^{\mu}_{i}}$, we see that the average energy at the final parameter point lies on the common tangent of both degenerate ground-state parity levels in the initial point.
Using Eq.~\eqref{eq:minimum} we obtain
\beq
\label{eq:finale}
\langle\epsilon_{f}\rangle\equiv\frac{\langle\hat{H}_{f}\rangle_{i}}{j}=\frac{1}{2}\left[\frac{h_{i}}{\lambda_{i}^{2}}\left(h_{i}\lambda_{f}-2h_{f}\lambda_{i}\right)-\lambda_{f}\right].
\eeq
This formula can be used as an approximation for the final energy average for a sufficiently large size of the system.
It allows us to estimate whether the dominant fraction of excitation at the final parameter values is located in the parity-broken phase, parity-restored phase, or in the transitional region around the ESQPT critical energy ${\epsilon_{c}=-h}$.
The approximate degeneracy of the initial ground-state parity doublet~\eqref{eq:initialstate} further implies that the first and second terms in Eq.~\eqref{eq:enevar} can be neglected, so the energy variance increases roughly quadratically with the length of the quench (the width of the final energy distribution grows linearly).
However, since the classical limit of the energy variance vanishes, any quantitative estimate of the width requires to account for finite-$j$ quantum effects. \\

\section{Results}\label{sec:results}
\subsection{Identification of zeros}

First, let us focus on the quenches that do not cross the ESQPT critical line at $\epsilon_{c}=-h$.
In Fig.~\ref{fig:Zpresentation} we present a backward quench ${\lambda_{i}=7}\to{\lambda_{f}=1.5}$ with equally weighted symmetry-broken eigenstates $\alpha=1/2$ and $\phi=0$. The probability associated to the resulting complexified partition function, $|\mathcal{Z}(\beta+it)|^{2}$, is represented in the $\beta\times t$ plane (left, middle); in particular, we represent the contours for several fixed values of such probability, i.e., the set of points $(\beta,t)$ such that $|\mathcal{Z}(\beta+it)|^{2}$ equals to a given constant. A pattern of zeros of $\mathcal{Z}(z)$ is clearly observed. The complexification of the time variable means that only zeros close to $\beta=0$ have a sensible impact on the survival probability, trademark of dynamical phase transitions. Yet, the curvature of the contours close to $\beta=0$ suggest that the survival probability is affected even if no zero falls exactly on the $\beta=0$ line. Therefore, these portraits provide insightful information on the behavior of the survival probability which, as we will argue later on, may be harder to identify merely by looking at the corresponding rate function. In the rightmost panel of Fig.~\ref{fig:Zpresentation} we represent the nodal lines of Eq.~\eqref{eq:Z}. Red and blue curves represent the null contour lines of the real and imaginary part of $\mathcal{Z}(\beta+it)$, respectively. The zeros observed in the middle panel neatly correspond to \textit{crossing points} of such nodal lines, where both real and imaginary parts of $\mathcal{Z}(\beta+it)$ vanish. Note that the contour lines of $\Re\,\mathcal{Z}(\beta,t)$ and $\Im\,\mathcal{Z}(\beta,t)$ always cross at the right angle, as can be analytically proven from the form of Eq.~\eqref{eq:Z}. The zeros of both survival probability and survival amplitude can either be deduced from the contour plots or calculated explicitly via the contour integral method explained below.

\subsection{Scaling with system size}

\begin{figure*}[t]
\centering
    \begin{tabular}{c c c}
\includegraphics[width=0.33\textwidth]{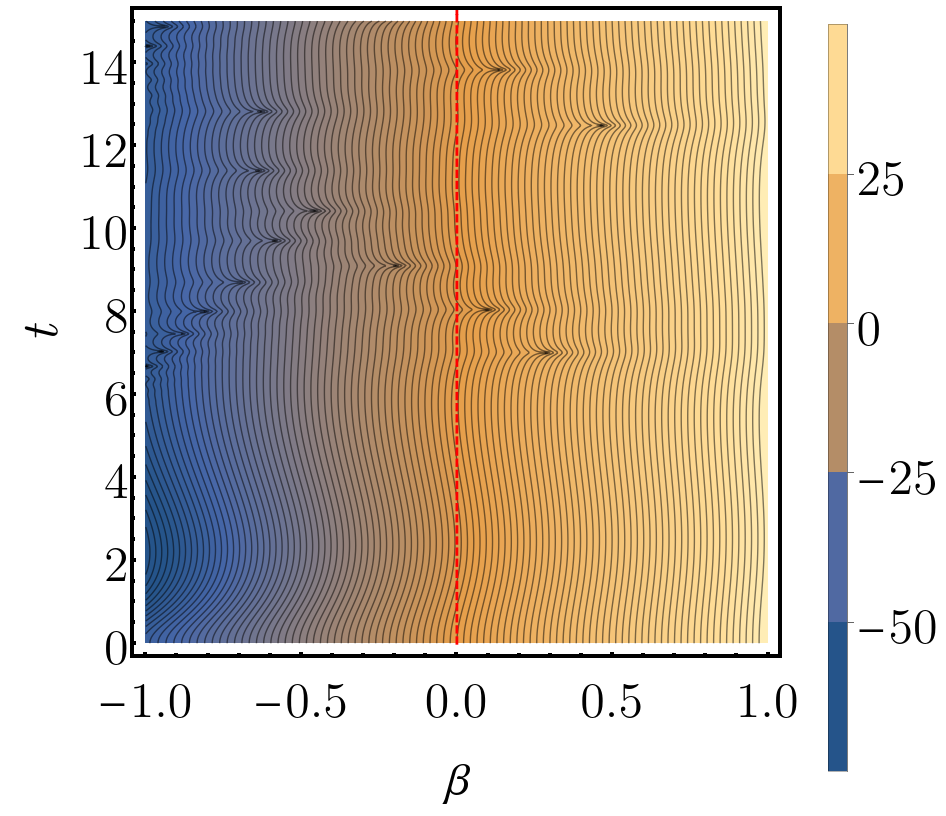} & \includegraphics[width=0.33\textwidth]{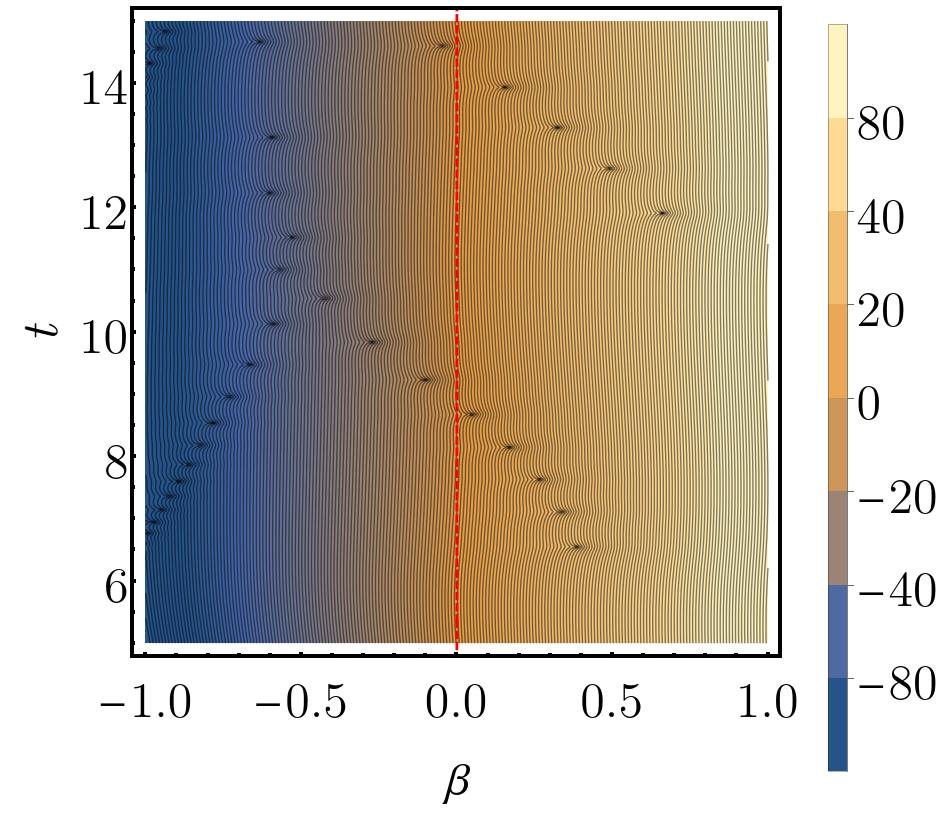} & \includegraphics[width=0.33\textwidth]{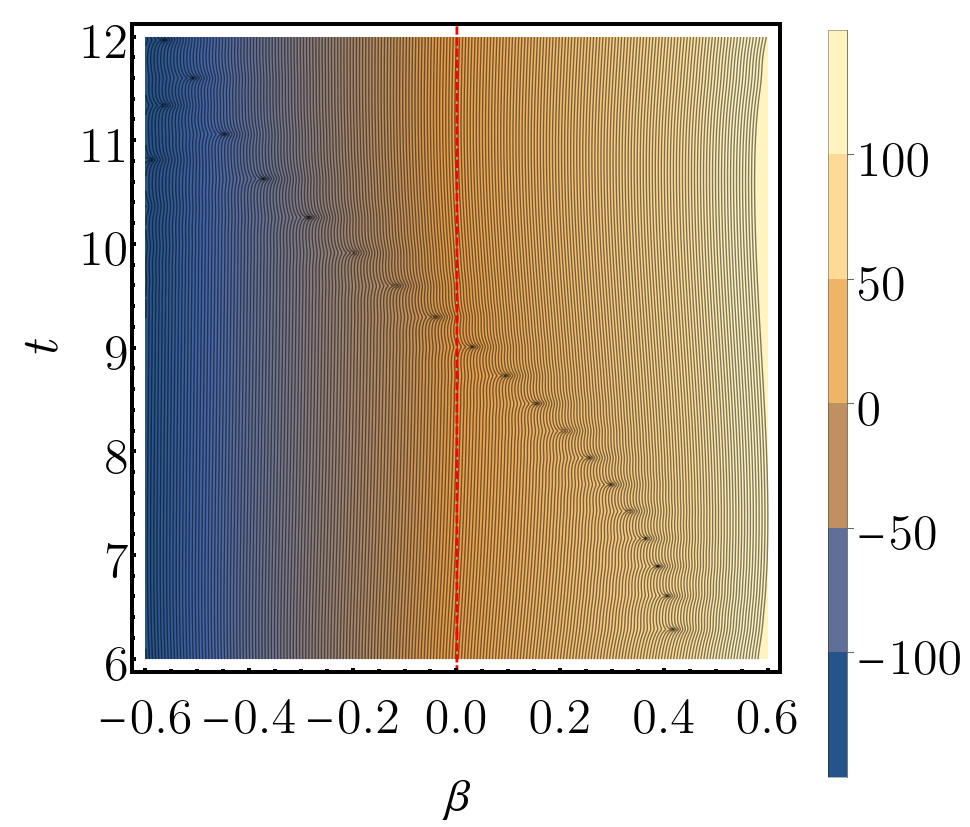}   \\

\hspace{-1cm} \includegraphics[width=0.33\textwidth]{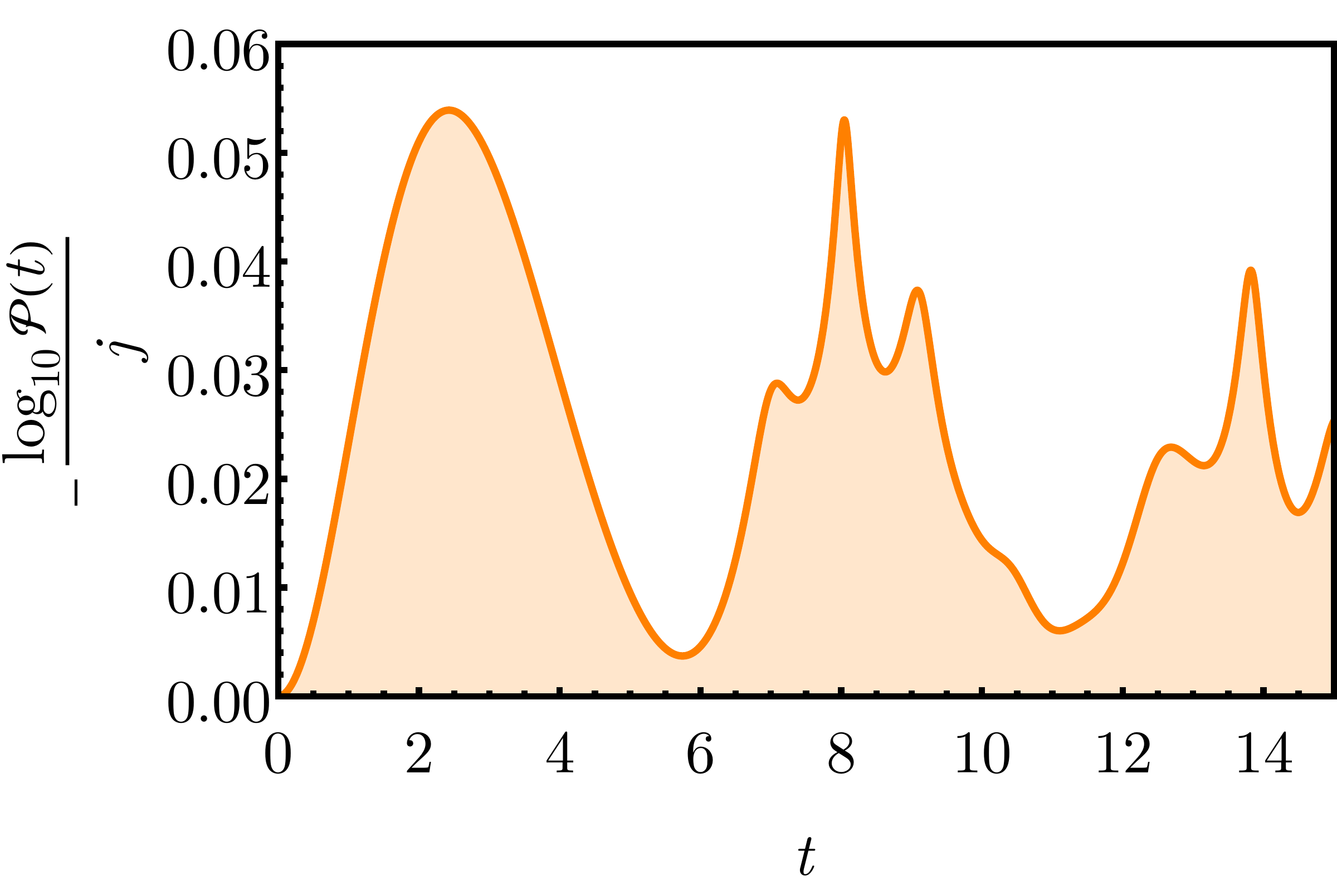} & \hspace{-1cm} \includegraphics[width=0.33\textwidth]{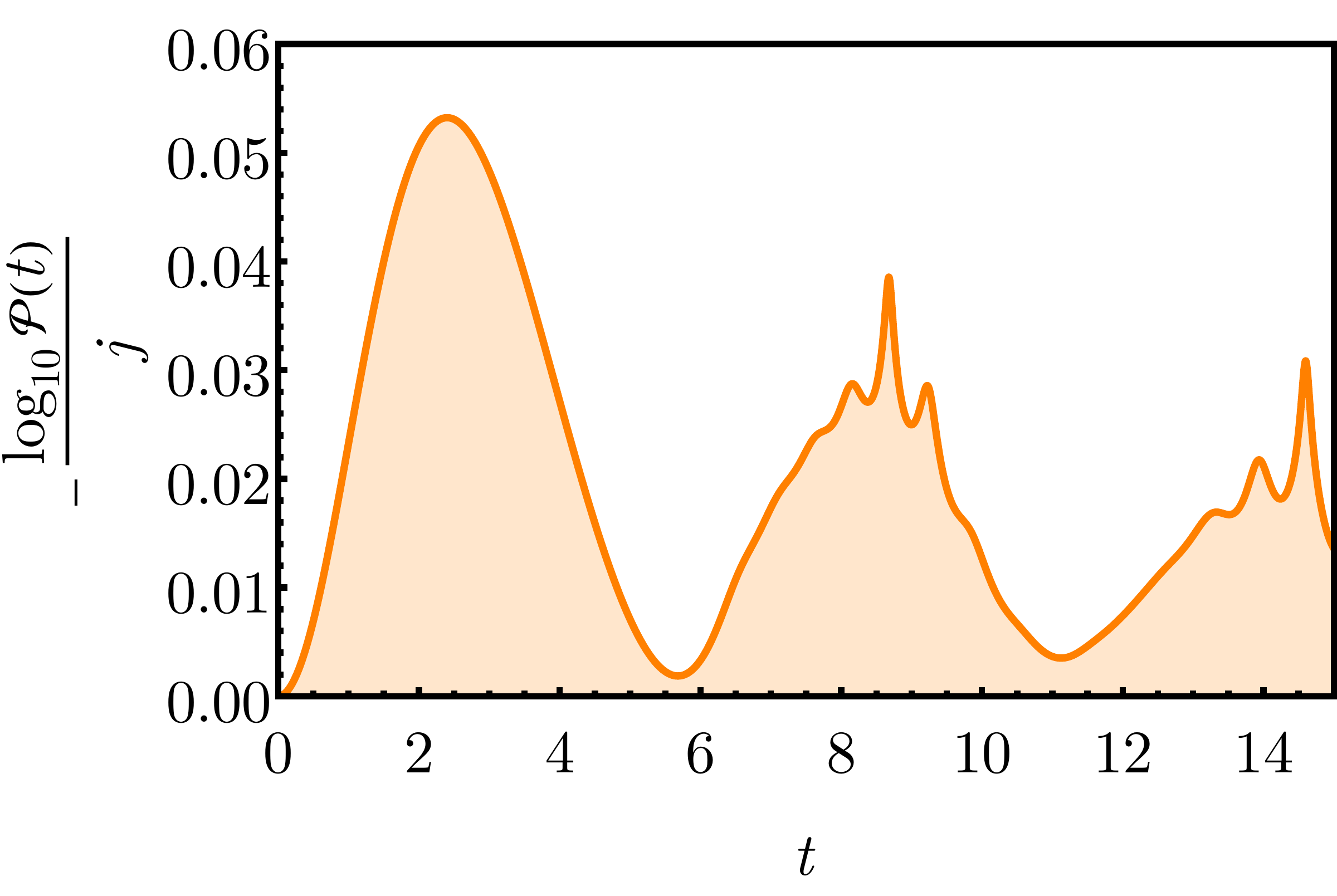} & \hspace{-1cm} \includegraphics[width=0.33\textwidth]{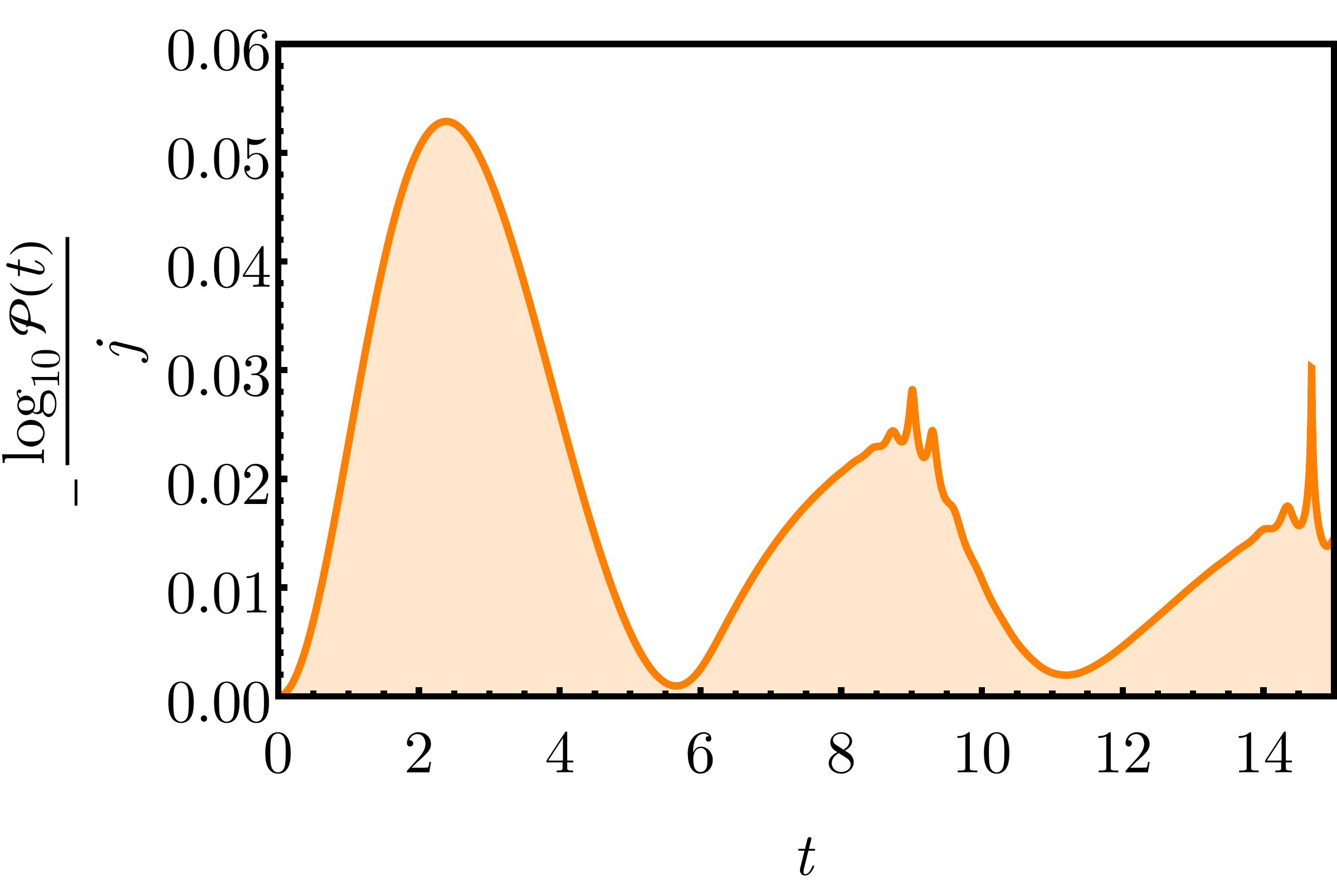} \\
    \end{tabular}
    \caption{Contour plots of $\log_{10}|\mathcal{Z}(\beta+it)|^{2}$ (top row) and corresponding rate functions of the survival probability, $\mathcal{P}(t)=|\mathcal{Z}(it)|^{2}$ (bottom row) for the quench ${\lambda_{i}=2.6}\to{\lambda_{f}=1.6}$, $\langle\epsilon_{f}\rangle=-1.07<\epsilon_{c}$. Vertical red dashed lines signal the $t$-axis ($\beta=0$).  System size is $j=50,100,200$ (left, middle, right columns), and the parameters in the initial state~\eqref{eq:initialstate} are $\alpha=1/2$ and $\phi=0$. Note that as the system size increases, a zoom in the contour plots is performed so as to improve visibility of a single group of zeros.}
    \label{fig:ZandSP}
\end{figure*}

In Fig.~\ref{fig:ZandSP} we study the zeros of the complex $\mathcal{Z}(\beta+it)$ after a relatively short backward quench with both $\lambda_{i}$ and $\lambda_{f}$ situated in the interacting phase $\lambda>\lambda_{c}$. Instead of analyzing the survival probability, $\mathcal{P}(t)$, itself, it is customary \cite{Heyl2013,Heyl2014} to analyze the rate function $r_{j}(t)=-\log_{10}\mathcal{P}(t)/j$.  We find a very good correspondence between the times when the rate function exhibits a kink and the times where a zero is close to the $\beta=0$ line in the complex-time partition function, namely, around $t\approx 8$ and $t\approx 14$. The first maximum of the rate function, around $t\approx 2$, does not correspond to a DPT because there is no zero around $\beta=0$ at $t\approx 2$ in $\mathcal{Z}(\beta+it)$; therefore, this maximum is of a different nature and will be discussed in more detail in Sec.~\ref{sec:IC}.

It is interesting to observe that the zeros of $\mathcal{Z}(\beta+it)$ appear to form ordered structures  similar to lines, and as the system size increases, so does the number of zeros within each structure, so as to form a continuum in the infinite-size limit. To corroborate this impression, we perform a finite-size scaling of the number of zeros of $\mathcal{Z}(\beta+it)$ within a fixed region of the $\beta\times t$ plane, and to count such number of zeros, $N_{0}=N_{0}(j)$, we proceed as described in what follows. The idea is to use the residue theorem of complex analysis, which teaches us that the integral of a complex function $f(z)$ around a closed path $\gamma$  is given by the sum of the residues of $f(z)$ at its poles. In a neighborhood of its zeros, $z_{0}$, the complexified partition function $\mathcal{Z}(\beta+it)$ may be expanded as 
\beq
\mathcal{Z}(z)\approx a_{1}(z-z_{0})+a_{2}(z-z_{0})^2+\mathcal{O}((z-z_0)^{3}),
\eeq
while its derivative behaves as 
\beq\label{eq:derZ}
\frac{d\mathcal{Z}(z)}{dz}\approx a_{1}+2a_{2}(z-z_{0})+\mathcal{O}((z-z_0)^{2}),
\eeq
with some $a_{k}\in\mathbb{C}$. Consider the following function: 
\beq
f(z)=\frac{d}{dz}\ln \mathcal{Z}(z).
\eeq
It may be shown that the zeros of $\mathcal{Z}(z)$ transform into poles of $f(z)$. Indeed, if $a_{1}\neq0$, then, close to a zero of $\mathcal{Z}(z)$,
\beq
f(z)\approx \frac{a_{-1}'}{z-z_{0}}+a_{0}'+a_{1}'(z-z_{0})+\mathcal{O}((z-z_0)^{2}),
\eeq
for some $a_{k}'\in\mathbb{C}$, in general $a_{k}'\neq a_{k}$. Importantly, $a_{-1}'=1$, which means that the residue of $f(z)$ at the pole is exactly unity, $\textrm{Res}_{z=z_{0}}f(z)=1$. If the root at $z=z_{0}$ happens to be of a higher order, i.e., if ${a_{1}=a_{2}=\dotsb=a_{k-1}=0}, {a_{k}\neq0}$, then the first $k$ terms in~\eqref{eq:derZ} vanish and the pole in $f(z)$ has $\textrm{Res}_{z=z_{0}}f(z)=k$, reflecting exactly the multiplicity of the root.
This enables us to use the residue theorem for straightforward calculation of the number
$N_{0}$ of roots of $\mathcal{Z}(z)$
for a fixed system size $j$,
\beq
\oint_{\gamma}f(z)dz=2\pi i N_{0}.
\eeq

\begin{figure}[t]
    \centering
\includegraphics[width=0.45\textwidth]{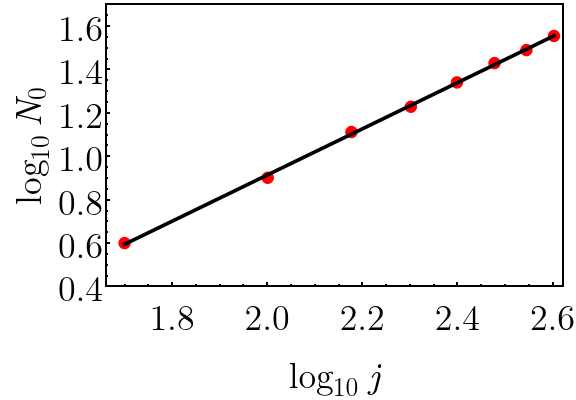} 
    \caption{Scaling of the number of zeros of $\mathcal{Z}(\beta+it)$ located in a rectangle centered around $\beta=0$, with $\beta\in(-0.5,0.5),t\in(6,12)$. A least-squares fit reveals the linear behavior $N_{0}(j)\sim j^{1.06335}$. The quench is $\lambda_{i}=2.6\to\lambda_{f}=1.6$, and the parameters in the initial state~\eqref{eq:initialstate} are $\alpha=1/2$ and $\phi=0$, as in Fig.~\ref{fig:ZandSP}.}
    \label{fig:scalingzeros}
\end{figure}

We have calculated $N_{0}$ for several system sizes $j$ in a small rectangular region of the complex plane containing the $\beta=0$ line, similar to the region depicted in the top, rightmost panel of Fig.~\ref{fig:ZandSP}. The result is shown in Fig.~\ref{fig:scalingzeros}, which reveals a strongly linear behavior $N_{0}\sim j$. This suggests that in the infinite-size limit, the density of zeros in the chains crossing the $t$ axis becomes infinite and the survival probability $\mathcal{P}(t)$ vanishes there.

\subsection{Dependence on the initial condition}\label{sec:IC}
It is easy to show that DPTs do not depend on the relative phase $\phi$ in a symmetry-broken initial state~\eqref{eq:initialstate}. For quantum systems with a parity symmetry, the complexified survival amplitude separates completely into its positive and negative parity contributions,
\beq\label{eq:Zalpha}
\begin{split}
\mathcal{Z}(z)&=\alpha\sum_{n}|c_{n,+}|^{2}e^{-zE_{n,+}(\lambda_{f})}\\&+(1-\alpha)\sum_{n}|c_{n,-}|^{2}e^{-zE_{n,-}(\lambda_{f})}
\end{split}
\eeq
where we have defined $c_{n,\pm}\equiv \bra{E_{n,\pm}(\lambda_{f})}\ket{E_{0,\pm}(\lambda_{i})}$ for simplicity. As a consequence, the survival probability, whose non-analytical points are used to define DPTs, do not depend on $\phi$ either, $\mathcal{P}(t)=|\mathcal{Z}(it)|^{2}$. 
Note, though, that the classical position of such initial state does depend on $\phi$ through
Eq.\,\eqref{eq:classic}.
The behavior of the survival probability, and thus that of the non-analytical times defining DPTs, in general depends on $\alpha$. 

\begin{figure}[h!]
    \centering

\begin{tabular}{c}
\hspace{-0.4cm}\includegraphics[width=0.36\textwidth]{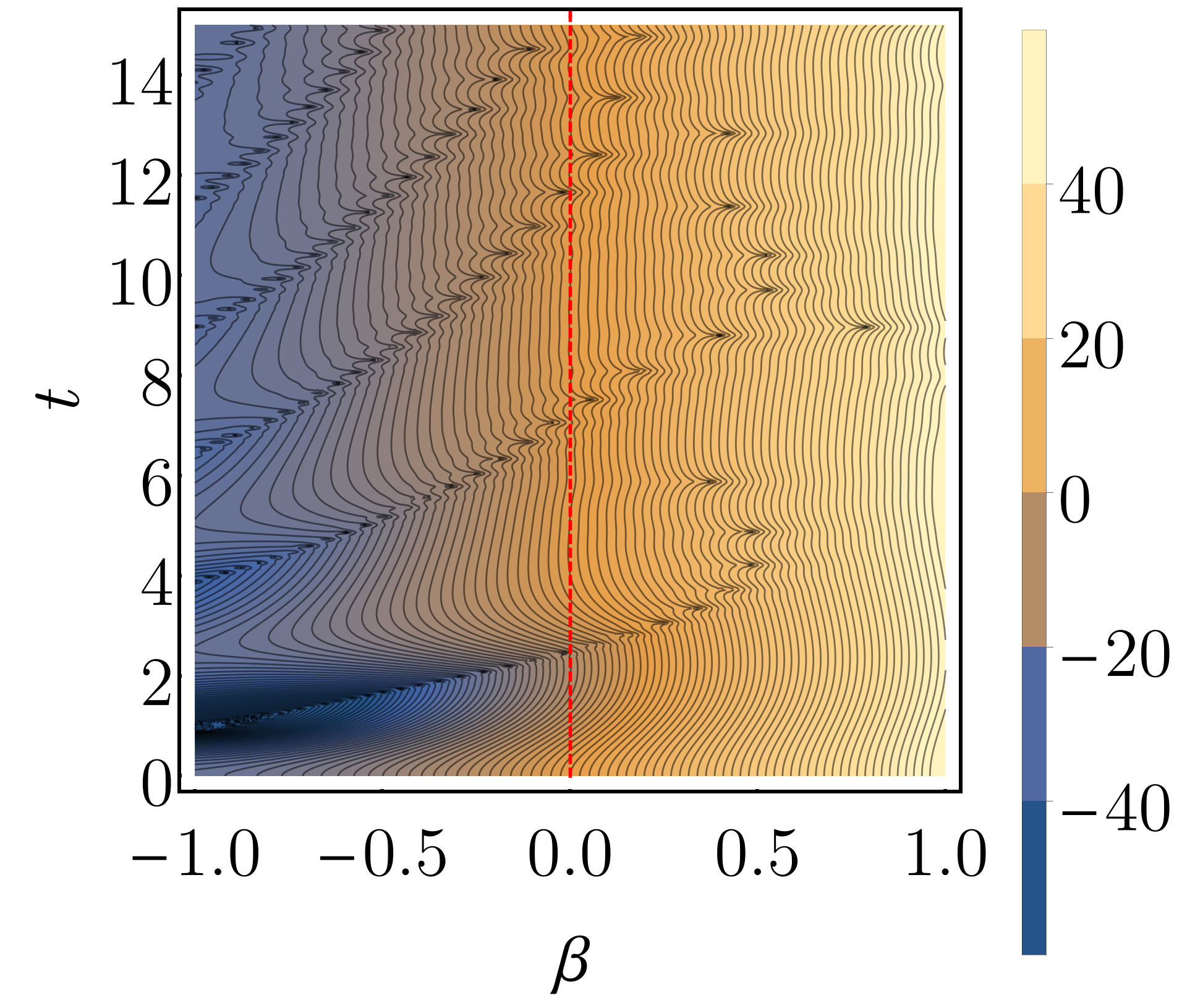} \\
\hspace{-0.2cm}\includegraphics[width=0.36\textwidth]{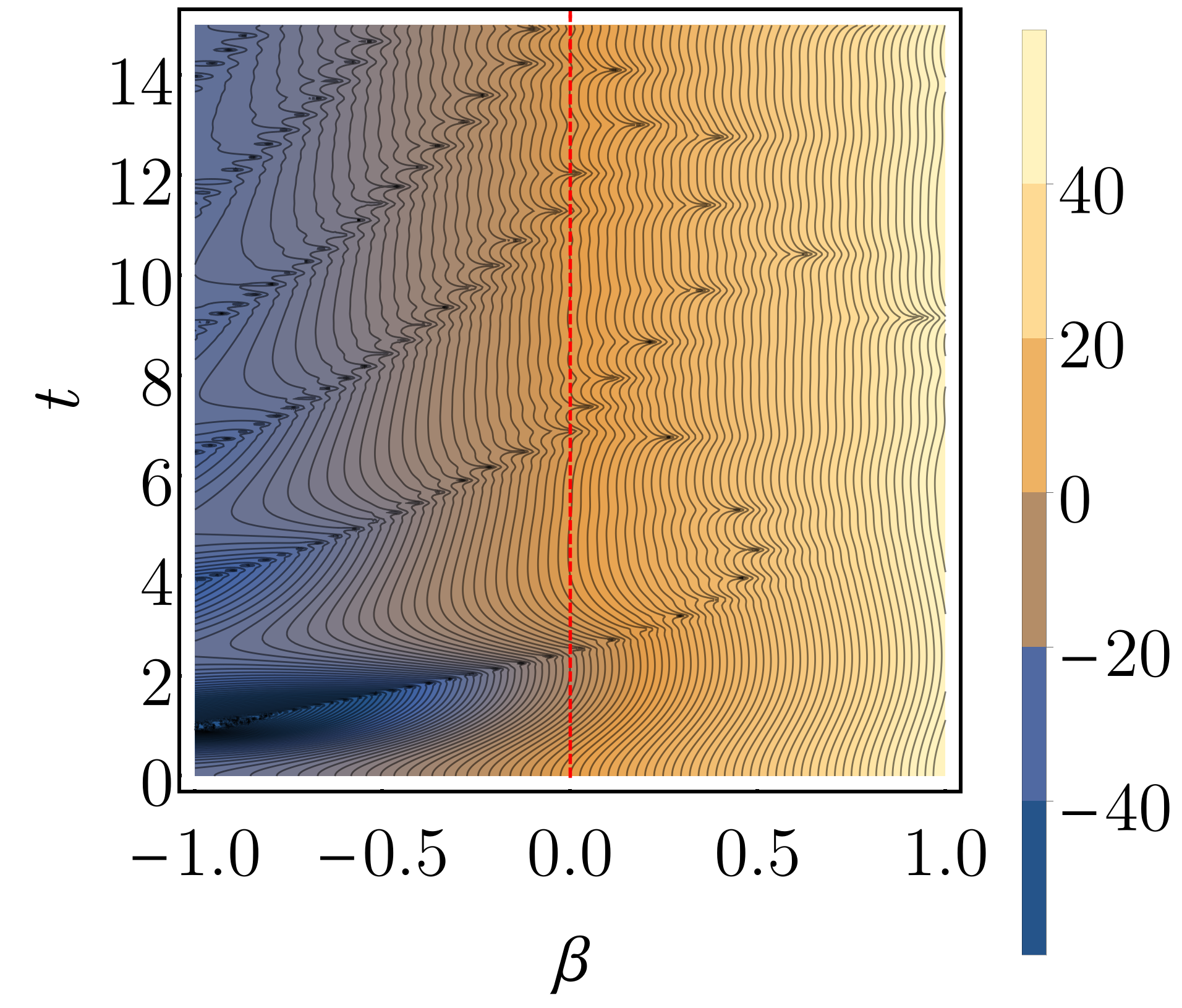}\\
\hspace{-1cm}\includegraphics[width=0.4\textwidth]{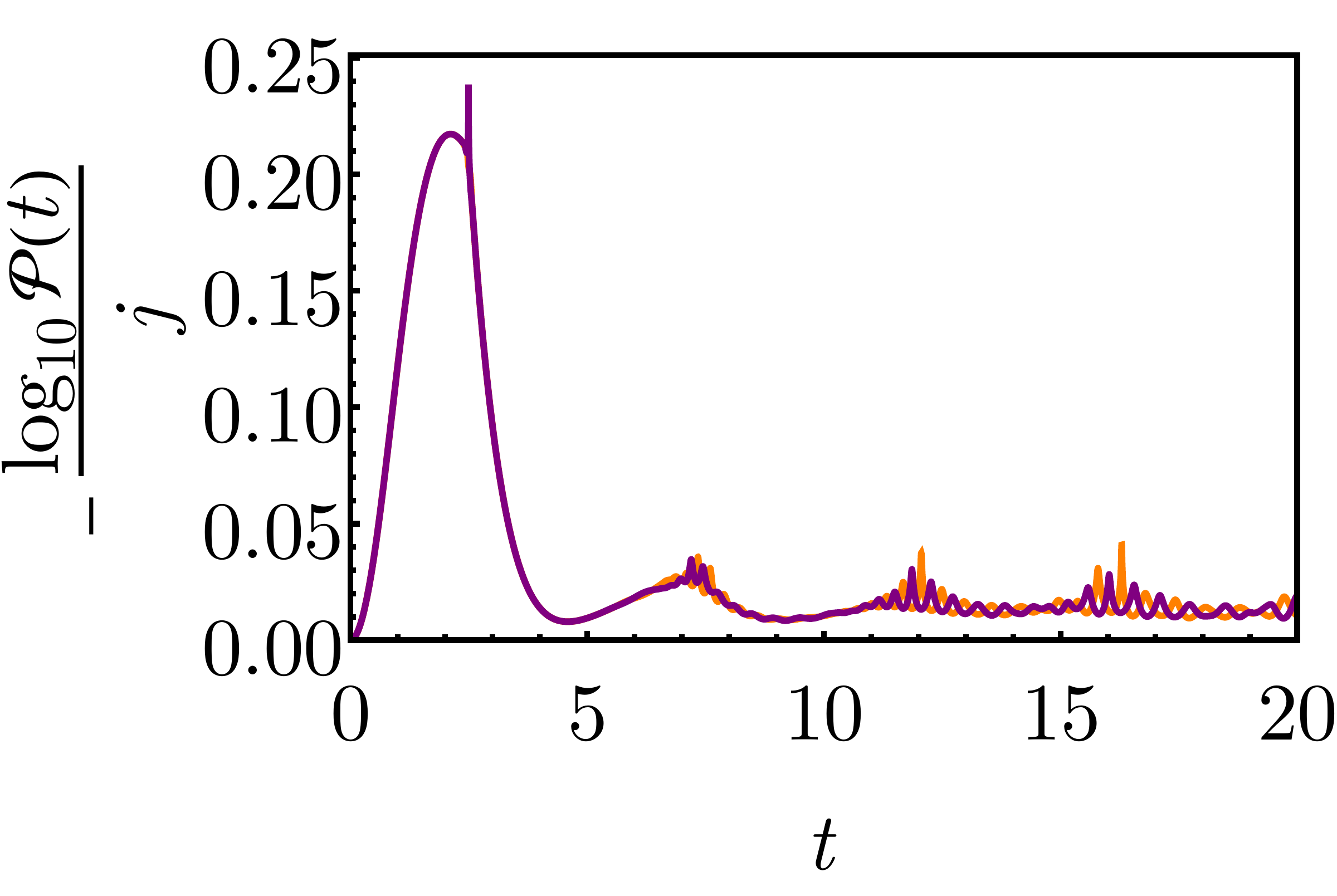}\\

    \end{tabular}
    \caption{Contour plot of $\log_{10}|\mathcal{Z}(\beta+it)|^{2}$ in the $(\beta,t)$ plane. Vertical red dashed lines signal the $t$-axis ($\beta=0$). The quench ${\lambda_{i}=10}\to{\lambda_{f}=1.6}$, $\langle\epsilon_{f}\rangle=-0.89>\epsilon_{c}$ has been performed.  The parameters in the initial state Eq. \eqref{eq:initialstate} are $\phi=0$, and $\alpha=0.1$ (top), $\alpha=0.9$ (middle). System size is $j=50$. Rate functions of the survival probability $\mathcal{P}(t)$ for the same quenches (bottom). Orange line corresponds to $\alpha=0.1$ and purple line to $\alpha=0.9$ ($j=100$). }
    \label{fig:alpha1minusalpha}
\end{figure}

If the quench only populates the eigenstates of the final Hamiltonian below the ESQPT critical line, then it was shown in \cite{Corps2022,Corps2022arxiv} that the populations of positive and negative states coincide up to a phase, that is, $|c_{n,+}|=|c_{n,-}|\equiv |c_{n}|$ in the infinite-size limit. This is a consequence of the exponentially vanishing gap between eigenlevels below the ESQPT, $E_{n,+}=E_{n,-}\equiv E_{n}$ if $E_{n,\pm}<E_{c}$, and also of the constancy of $\hat{\mathcal{C}}$ \cite{Corps2021}. Therefore, \textit{in the infinite-size limit}, it follows from Eq.~\eqref{eq:Zalpha} that $\mathcal{Z}(z)$ is also independent of $\alpha$ as these contributions cancel out. Hence $\mathcal{Z}(z)=\sum_{n}|c_{n}|^{2}e^{-zE_{n}}$ as in Eq.~\eqref{eq:Z}.
For finite-size systems, the above equality is to be taken approximately; however, the exponentially close energy doublets and the constancy of $\hat{\mathcal{C}}$ as $j$ increases \cite{Corps2021} guarantee that finite-size effects are negligible already for quite small values of $j$. 

If, however, the quench only populates eigenstates of the final Hamiltonian above the ESQPT critical line, where there are no parity quasidublets, $E_{n,+}\neq E_{n,-}$, and $\hat{\mathcal{C}}$ does not behave as a constant in the infinite-size limit, $|c_{n,+}|\neq |c_{n,-}|$, the previous argument does not hold true and the dependence of $\mathcal{Z}(z)$ on $\alpha$ remains even in the large-$j$ limit. Since the negative-parity eigenlevels are systematically larger than the positive-parity ones, $E_{n,-}>E_{n,+}$ (c.f. Fig.~\ref{fig:lipkinphase}), the positive and negative parity contributions to Eq.~\eqref{eq:Zalpha} are not on the same footing, and therefore $\mathcal{Z}(z)$ is not symmetric under $\alpha\to 1-\alpha$. This can be observed in Fig.~\ref{fig:alpha1minusalpha}, where $|\mathcal{Z}(z)|^{2}$ is represented for a quench departing from two initial states with $\alpha=0.1$ or $\alpha=0.9$. The exact position of zeros is different in each case, as expected; nevertheless, the zeros at the $\beta=0$ line occur more or less at equivalent times, meaning that the kinks in the corresponding rate functions are very close for both initial states. This is clearly illustrated in the bottom panel of Fig.~\ref{fig:alpha1minusalpha}. Although the rate functions for $\alpha=0.1$ and $\alpha=0.9$ are not the same, they are influenced by the zeros of $\mathcal{Z}(\beta+it)$ at similar times.

\begin{figure}[h!]
    \centering

\begin{tabular}{cc}
\hspace{-2.8cm}\includegraphics[width=0.25\textwidth]{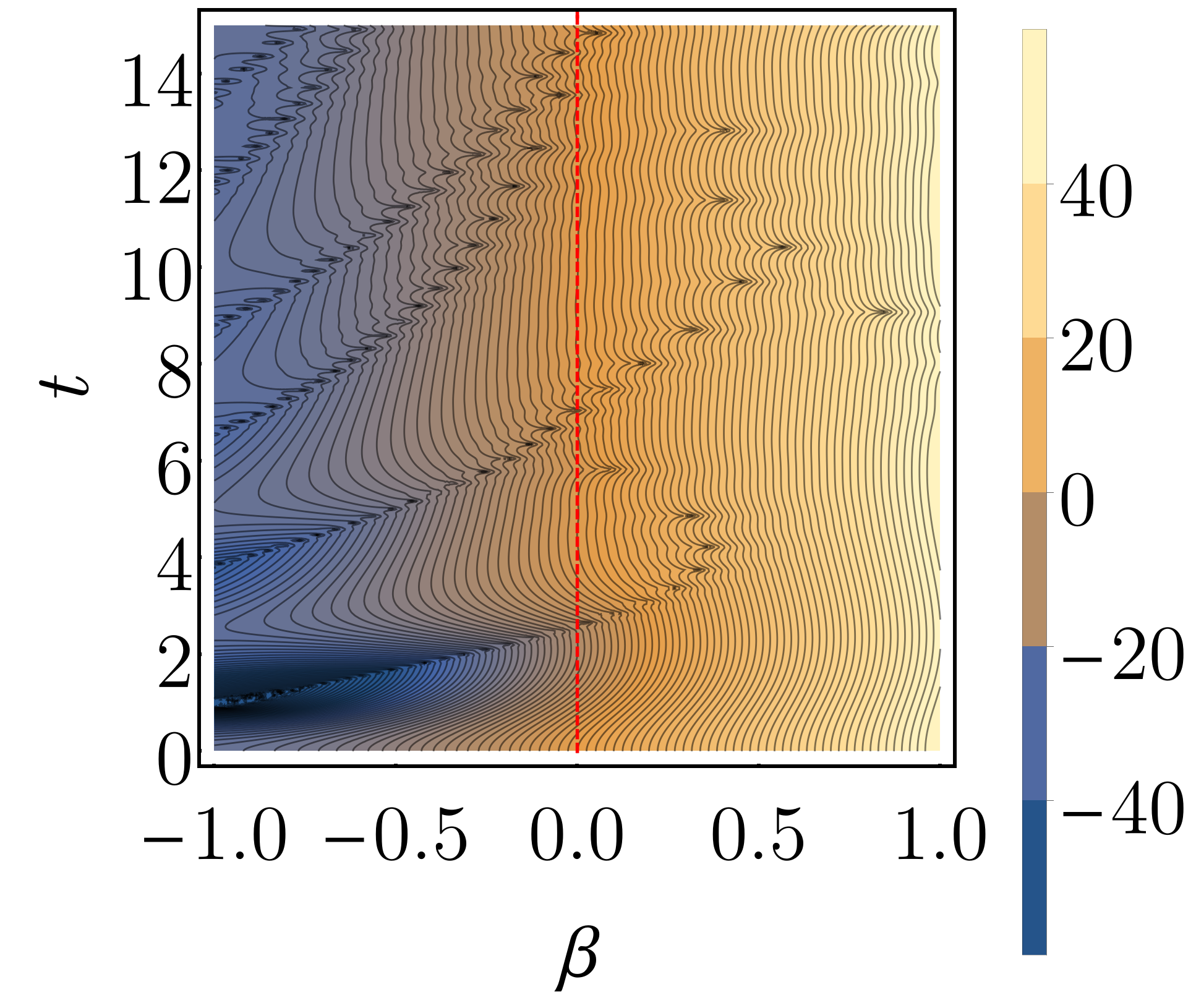} & 
\hspace{-2.3cm}\includegraphics[width=0.25\textwidth]{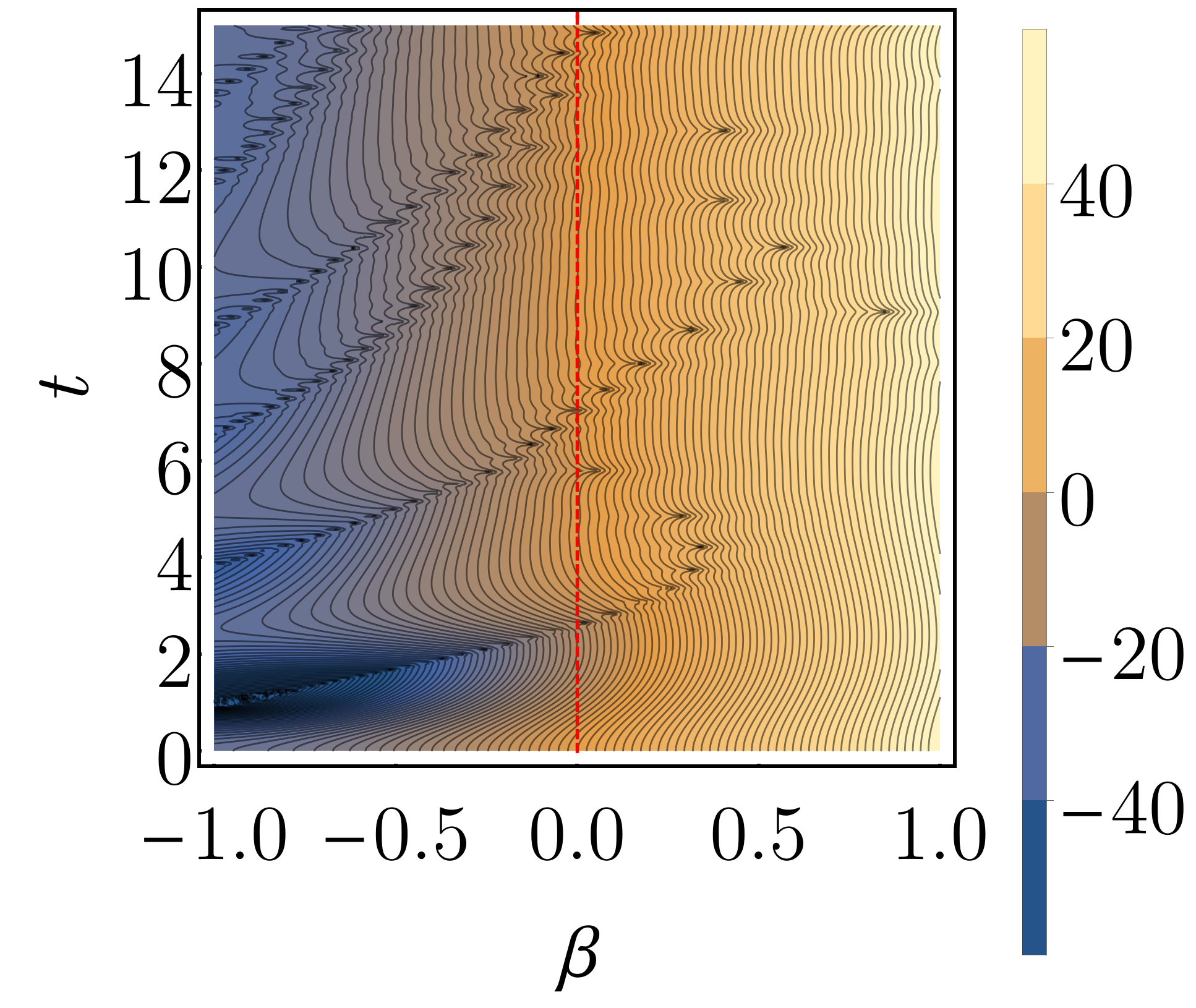}\\

\hspace{-2.8cm}\includegraphics[width=0.25\textwidth]{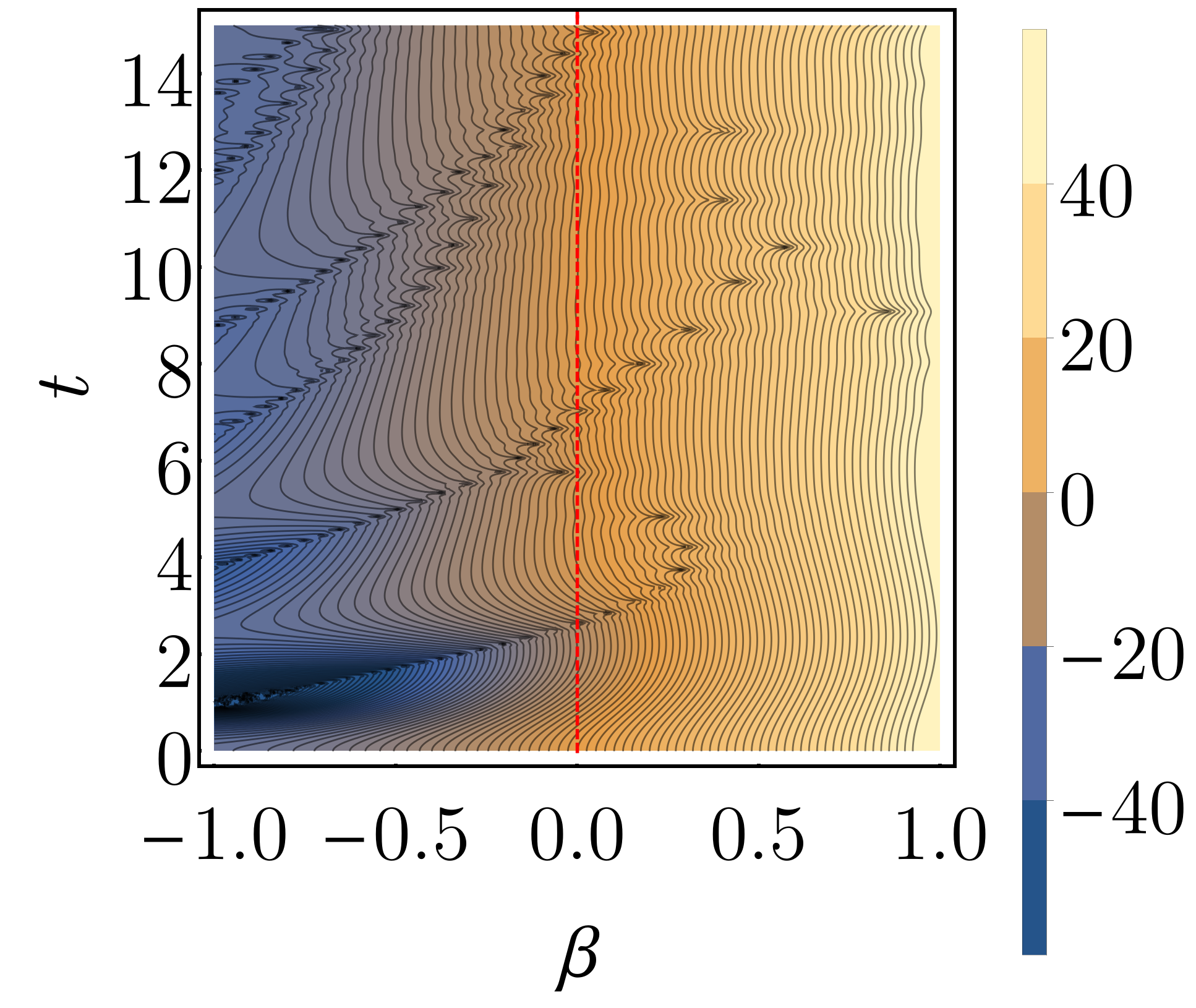} & 
\hspace{-2.3cm}\includegraphics[width=0.25\textwidth]{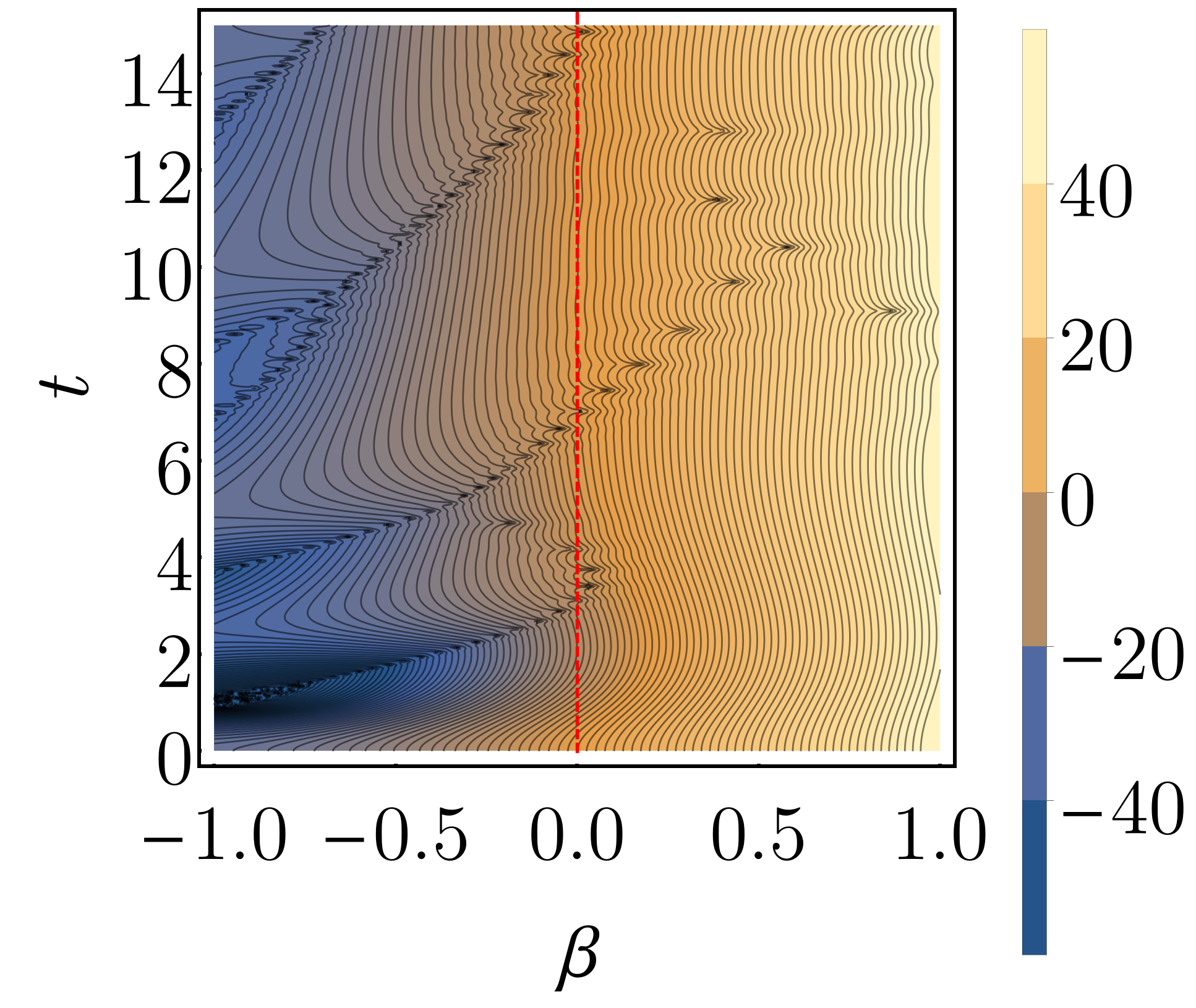} \\

\hspace{-2.8cm}\includegraphics[width=0.25\textwidth]{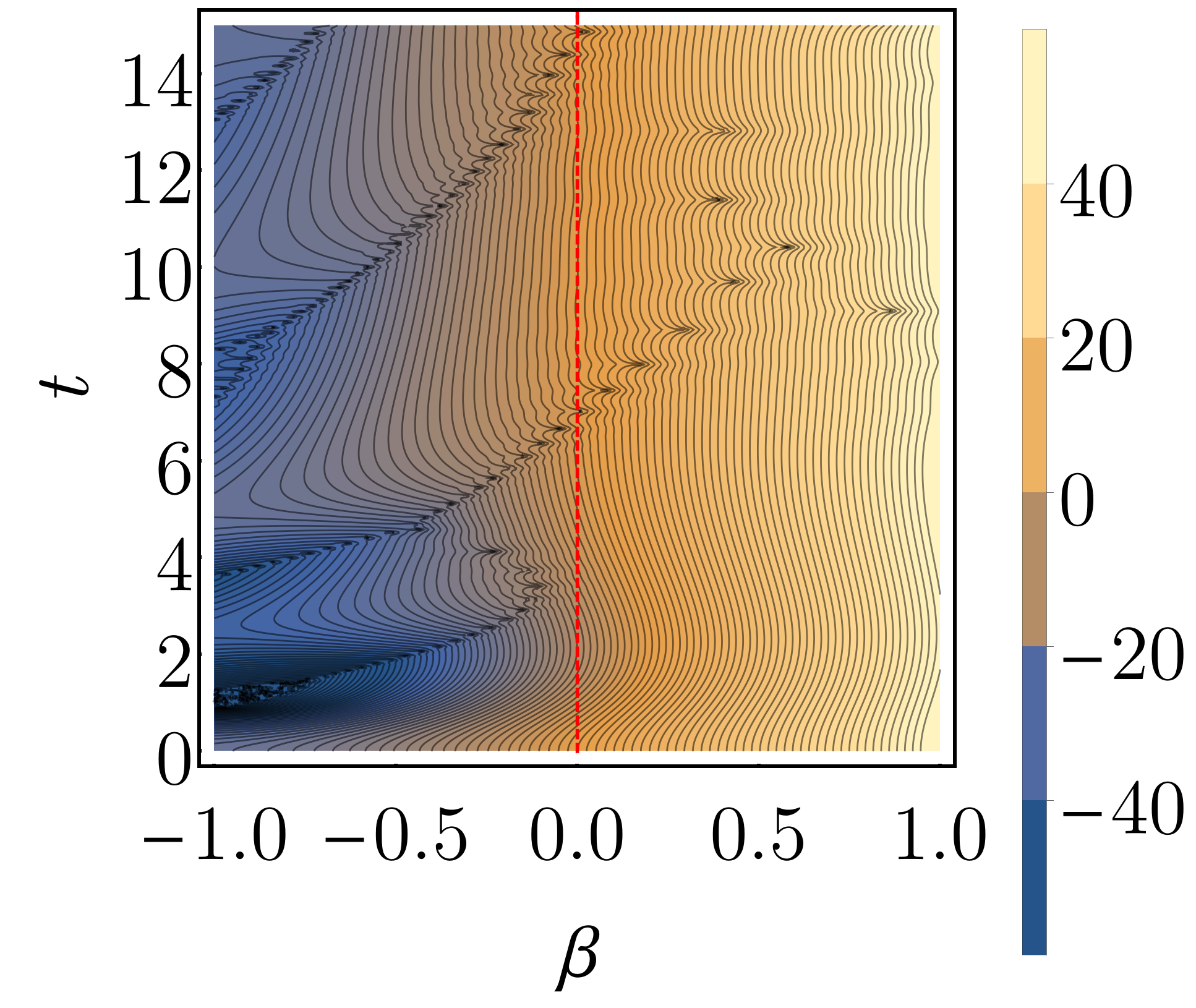} & 
\hspace{-2.3cm}\includegraphics[width=0.25\textwidth]{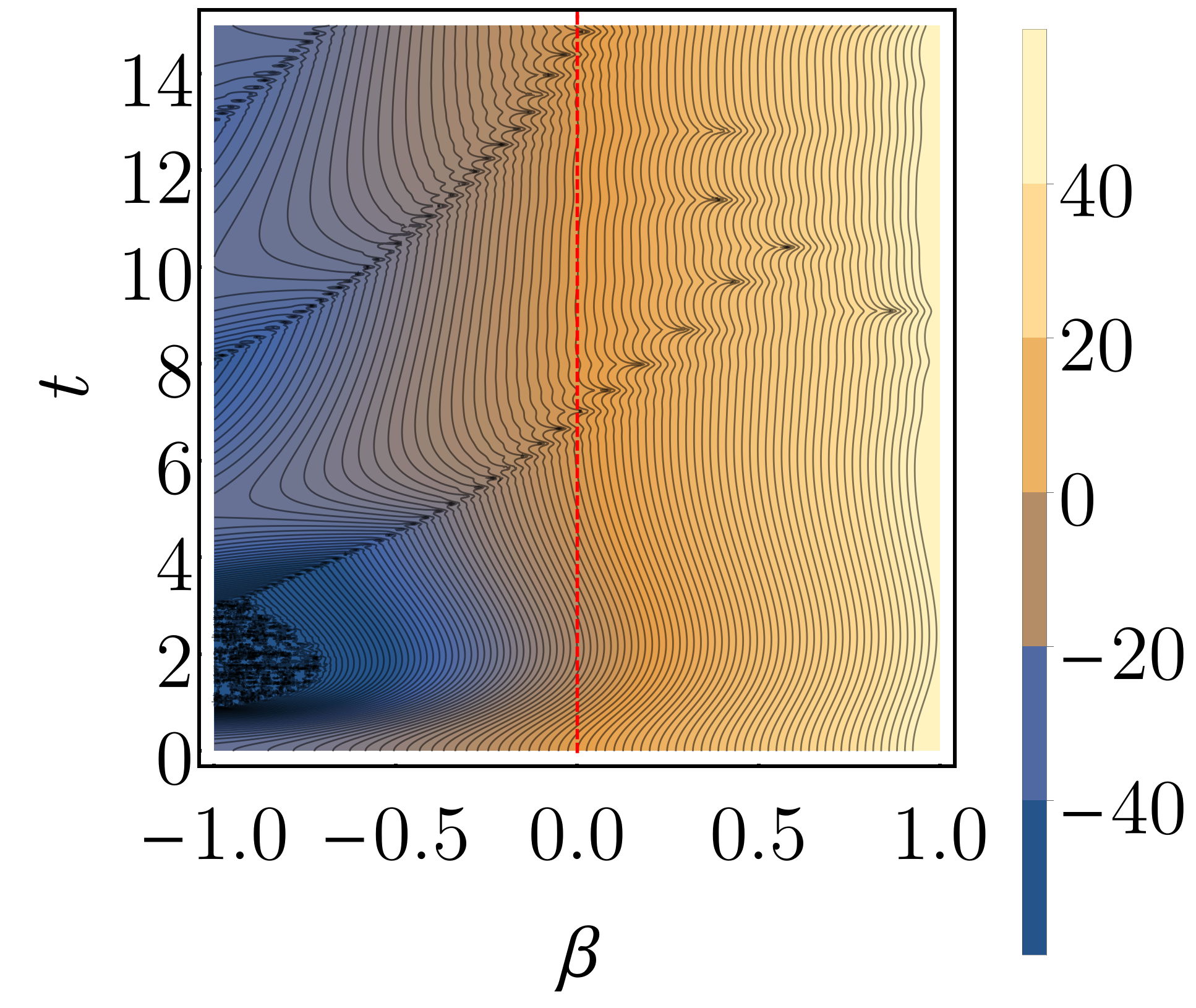}\\

    \begin{tabular}{c}
\hspace{0cm}\includegraphics[width=0.35\textwidth]{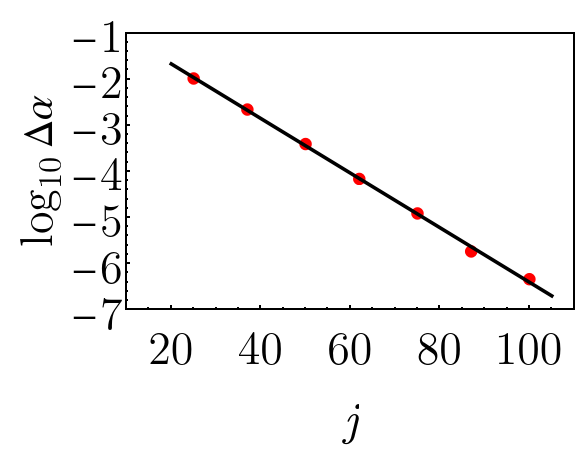} \\
\end{tabular}
    \end{tabular}
    \caption{Contour plot of $\log_{10}|\mathcal{Z}(\beta+it)|^{2}$ in the $(\beta,t)$ plane. Vertical red dashed lines signal the $t$-axis ($\beta=0$). The quench ${\lambda_{i}=10}\to{\lambda_{f}=1.6}$ has been performed.  The parameters in the initial state~\eqref{eq:initialstate} are $\phi=0$, and $\alpha=0.4,0.42$ (top row, from left to right), $\alpha=0.45,0.499$ (middle row), and $\alpha=0.49995,0.5$ (bottom row). System size is $j=50$. Scaling of $\Delta\alpha(j) \equiv |\alpha_{c}(j)-\alpha_{c}(\infty)|$, $\alpha_{c}(\infty)=0.5$, with system size, revealing an exponential decay $\Delta \alpha(j) \sim 10^{aj}$ with $a\approx -0.059$. }
    \label{fig:scalingalpha}
\end{figure}

To end this section, we will highlight an interesting effect taking place when the initial state is built with an even proportion of positive and negative parity eigenstates, $\alpha=0.5$, which is the initial state commonly prepared in experiments~\cite{Muniz2020}. {If $\phi=0$, this state has $\langle \hat{\mathcal{C}}\rangle=1$. Therefore, it is fully localized in one of the two wells, see Fig.~\ref{fig:lipkinphase}, and converges to a classical state in the infinite-size limit. Any other initial state with $\alpha\neq 0.5$ is a superposition of left and right classical wells, with $\langle \hat{\mathcal{C}}\rangle\in[0,1)$. In Fig.~\ref{fig:scalingalpha} we have represented contour plots of $|\mathcal{Z}(\beta+it)|^{2}$ for a quench ${\lambda_{i}=10}\to{\lambda_{f}=1.6}$ from the initial states~\eqref{eq:initialstate} with $\phi=0$ and varying values $\alpha\in[0.4,0.5]$ in order to study the dynamics of the zeros as $\alpha=0.5$ is slowly approached. It is clearly observed that as $\alpha$ increases, the upper lines of zeros seem to collapse in pairs, revealing a zipper-like structure which gradually closes from higher to lower values of $\beta$ with increasing $\alpha$.
At $\alpha=0.5$, the previously separated four chains of zeros present for lower values of $\alpha$ have completely merged into two separate lines. This shows that the distribution of non-analytical points in the survival probability explicitly depends on $\alpha$, i.e., on the fine details of the initial state, not just on its energy or the initial value of the control parameter.

The two chains of zeros starting from the bottom at $\alpha=0.4$ display an even more peculiar behavior. They also eventually merge into a single one at $\alpha=0.5$, but instead of doing so in the zipper-structure way, the first chain of zeros bends backwards and gradually recedes towards the lower values of $\beta$. Importantly, there is some value of $\alpha$ below which the first non-analytical point in the survival probability, near $t=2$, completely disappears as the structure of zeros moves to the left of $\beta=0$, leaving no trace in $\mathcal{P}(t)=|\mathcal{Z}(0+it)|^{2}$.  This induces structural changes in the positions of the kinks in the rate functions. The terms \textit{regular} and \textit{anomalous} dynamical phase transitions have been used to refer to the time when the first non-analytical point occurs in the survival probability; it is called regular if $\mathcal{P}(t)$ shows its first nonanalyticity at the first revival, and anomalous otherwise~\cite{Homrighausen2017}. In our system, the case $\alpha=0.4$ corresponds to a \textit{regular} DPT, while $\alpha=0.5$ results in an \textit{anomalous} DPT instead. 

The critical value of $\alpha$ separating both scenarios depends on system size, $\alpha_{c}(j)$, and increases with it. To analyze the finite-size precursor $\alpha_{c}(j)$ at which the bottom line of zeros has crossed the $\beta=0$ line to the left, we employ a bisection algorithm. The procedure is as follows: in the first iteration, we start from $\alpha_{L}=0.48$ and $\alpha_{R}=0.5$, and calculate the middle point $\alpha_{M}=(\alpha_{L}+\alpha_{R})/2=0.49$. We choose the two values of $\alpha$ between which there is a change in the position of the zeros, from being to the left of $\beta=0$ to crossing the $\beta=0$ line, and these values of $\alpha$ are used to iterate the process. We have followed this procedure for several values of the size parameter $j$, applying twenty iterations of the described algorithm. The results are shown in the bottom panel of Fig.~\ref{fig:scalingalpha}. We obtain the scaling behavior $\Delta\alpha(j)\sim 10^{aj}$ with $a\approx -0.059$, where $\Delta\alpha(j)\equiv |\alpha_{c}(j)-\alpha_{c}(\infty)|$ is the difference of the finite-size precursor $\alpha_{c}(j)$ and its value in the $j\to\infty$ limit, $\alpha_{c}(\infty)$. Choosing $\alpha_{c}(\infty)=0.5$, denoting a perfectly evenly symmetry-broken initial state, yields clear results of an exponentially diminishing size of the anomalous DPT region.

The results of this section illustrate the ways
in which the initial state influences the appearance of DPTs in the survival probability.
We saw, in particular, that these structures depend on whether the quenched state ends in a symmetry-broken or symmetry-restored phase, and also on the particular form of the parity-broken initial state.

\subsection{Dependence on the energy of the quenched state}

The complex-time survival amplitude~\eqref{eq:Z} contains dynamical information directly related to the quench performed. For quench protocols $\lambda_{i}\to\lambda_{f}$, the population of the eigenstates of the final Hamiltonian is non-universal and strongly sensitive to the specific choice of quench parameters, as revealed by the LDOS (see Sec.\,\ref{sec:qu}). For the infinite-range transverse-field Ising model, the parity symmetry means that the full LDOS, $P(E)$, separates into two parts,
\begin{align}
&P(E)=P_{+}(E)+P_{-}(E)\\&=\sum_{n}|c_{n,+}|^{2}\delta(E-E_{n,+})+\sum_{n}|c_{n,-}|^{2}\delta(E-E_{n,-}),\nonumber
\end{align}
where $c_{n,\pm}\equiv \bra{E_{n,\pm}(\lambda_{f})}\ket{\Psi_{0}(\lambda_{i})}$, with $P_{\pm}(E)$ denoting the population of positive/negative final eigenstates. 
The initial state $\ket{\Psi_{0}(\lambda_{i})}$ is generally not an eigenstate of the final Hamiltonian 
and its average energy is given by Eq.~\eqref{eq:eneav}.
It is well-known that the critical times and the shape of the survival probability depend on to which part of the energy spectrum the initial state is led by the quench~\cite{Corps2022,Kloc2018}. In the context of dynamical phase transitions, phase diagrams have been determined on the basis of such changes. Here we study the effect that the driving of the final average energy of the quench has on the complex-time survival amplitude.

\begin{figure*}[t]
\centering
    \begin{tabular}{c c c}
\includegraphics[width=0.33\textwidth]{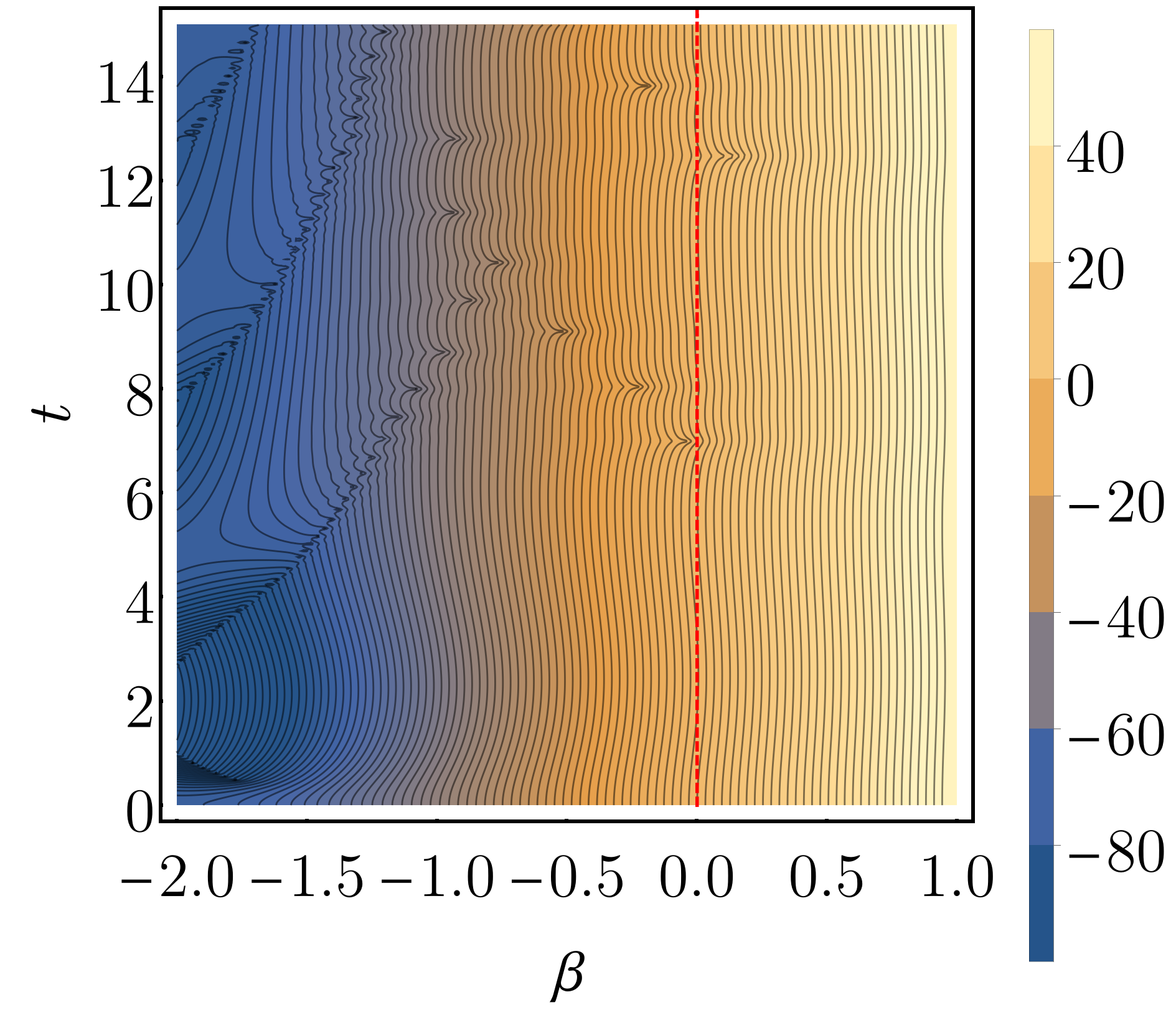} & \includegraphics[width=0.33\textwidth]{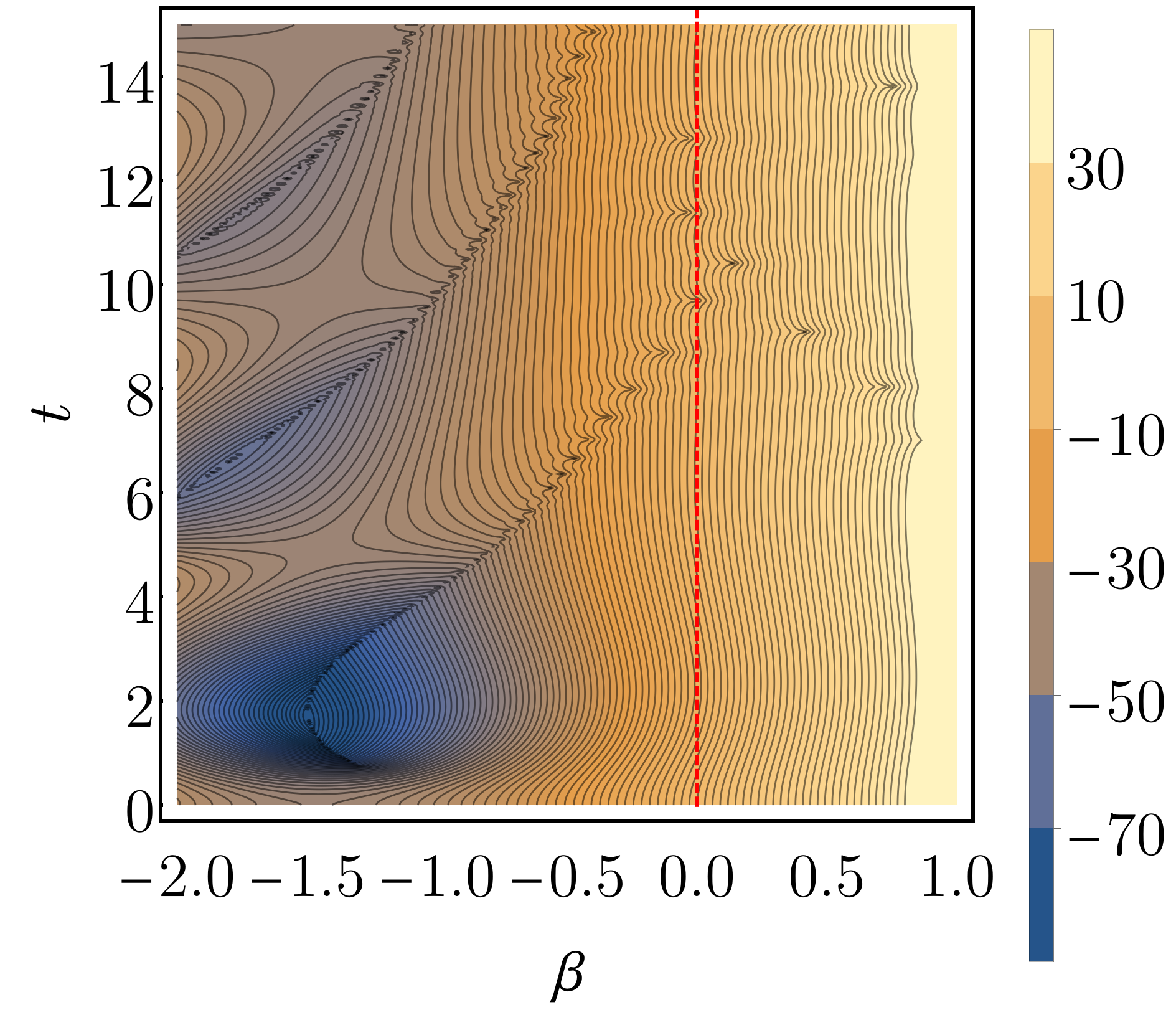} & \includegraphics[width=0.33\textwidth]{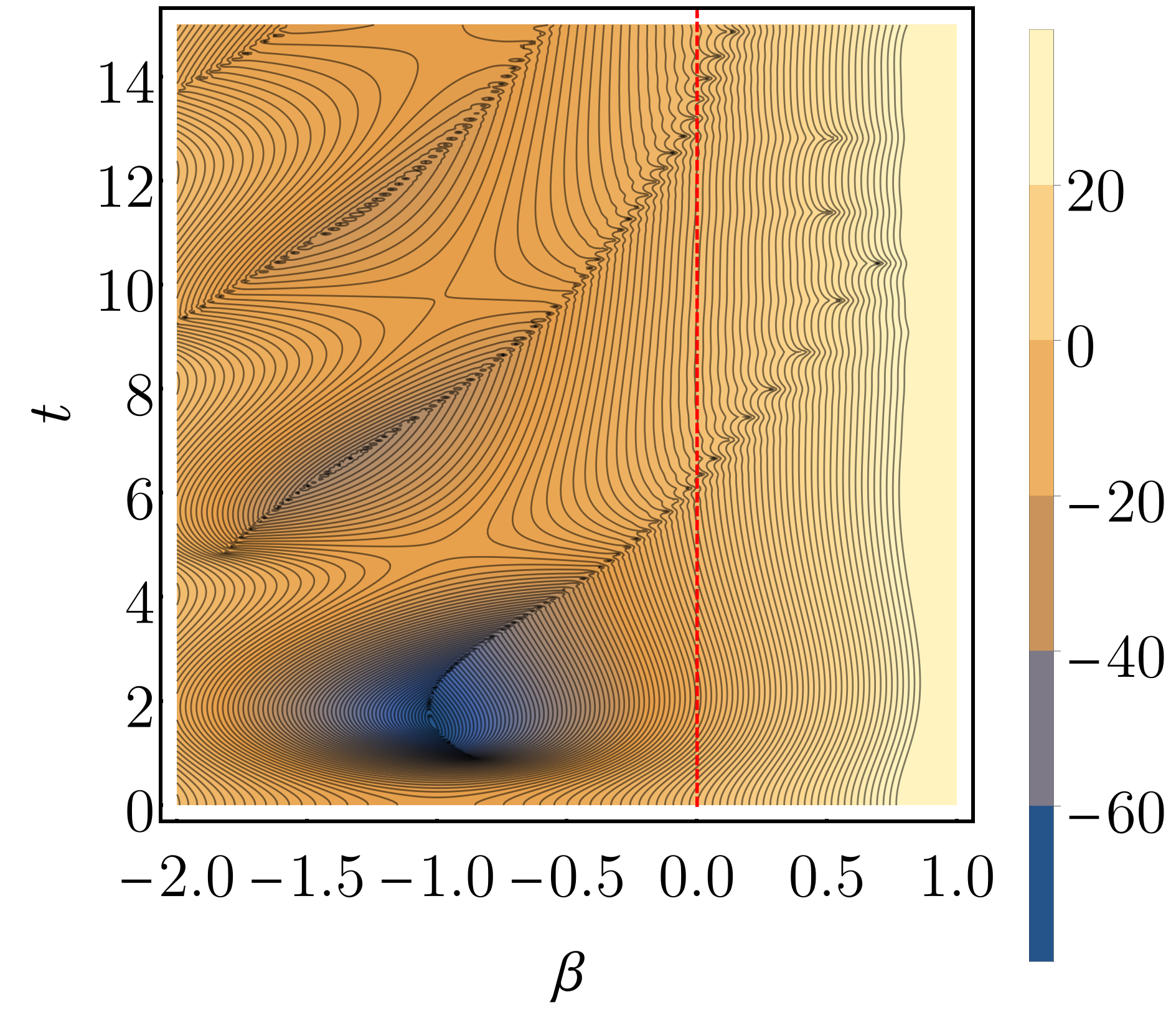}   \\

\hspace{-1cm} \includegraphics[width=0.33\textwidth]{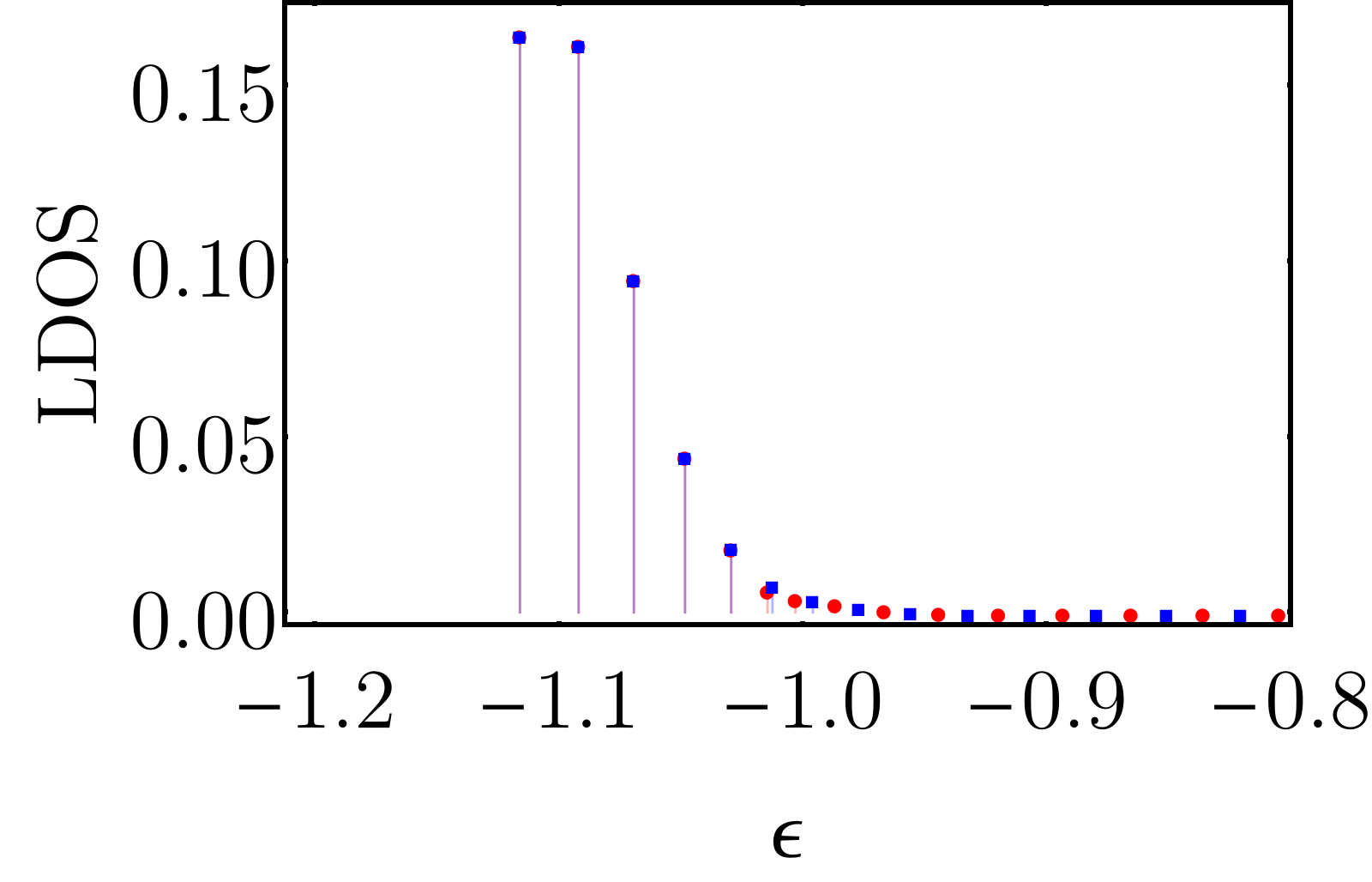} & \hspace{-1cm} \includegraphics[width=0.33\textwidth]{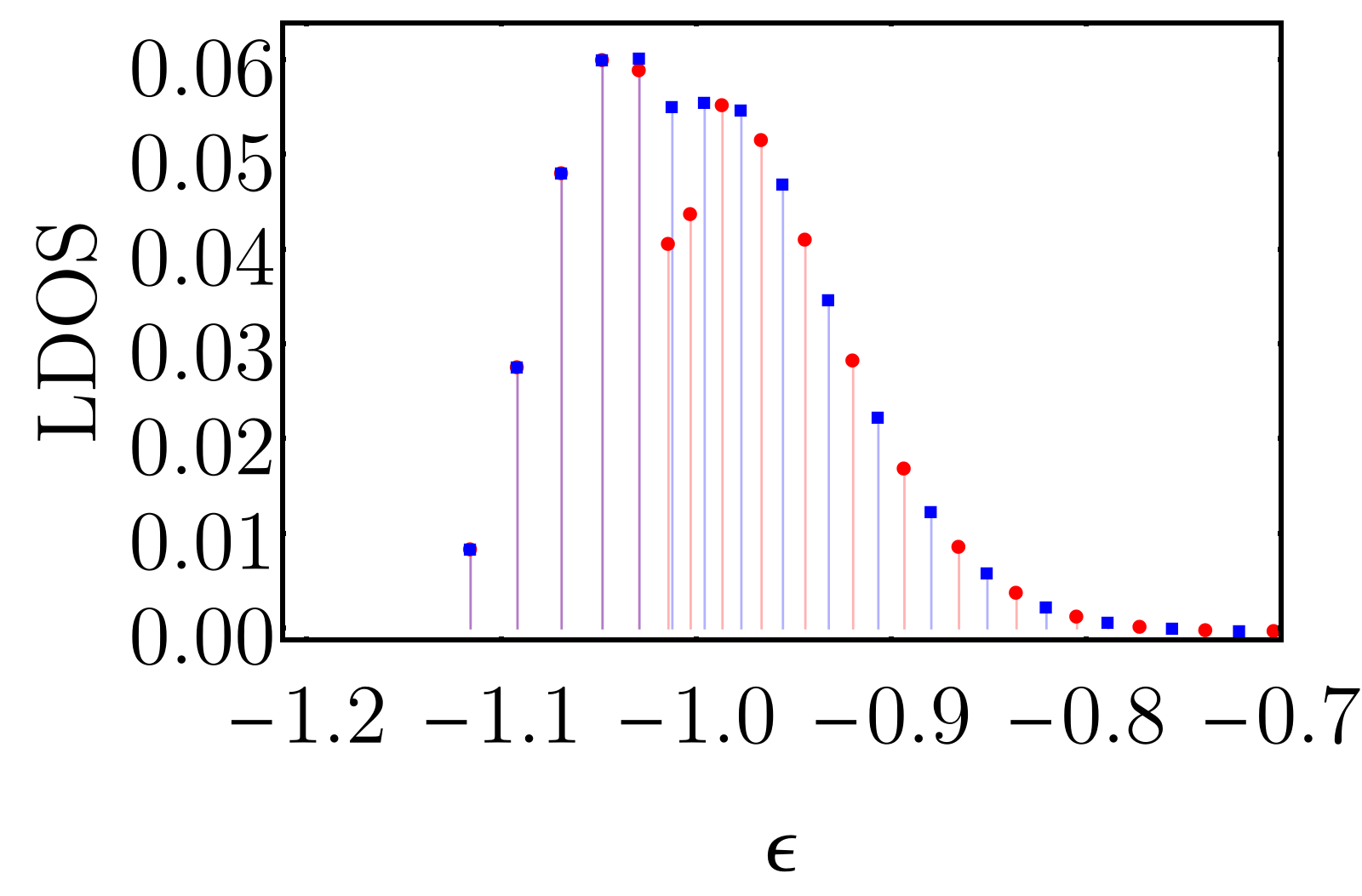} & \hspace{-1cm} \includegraphics[width=0.33\textwidth]{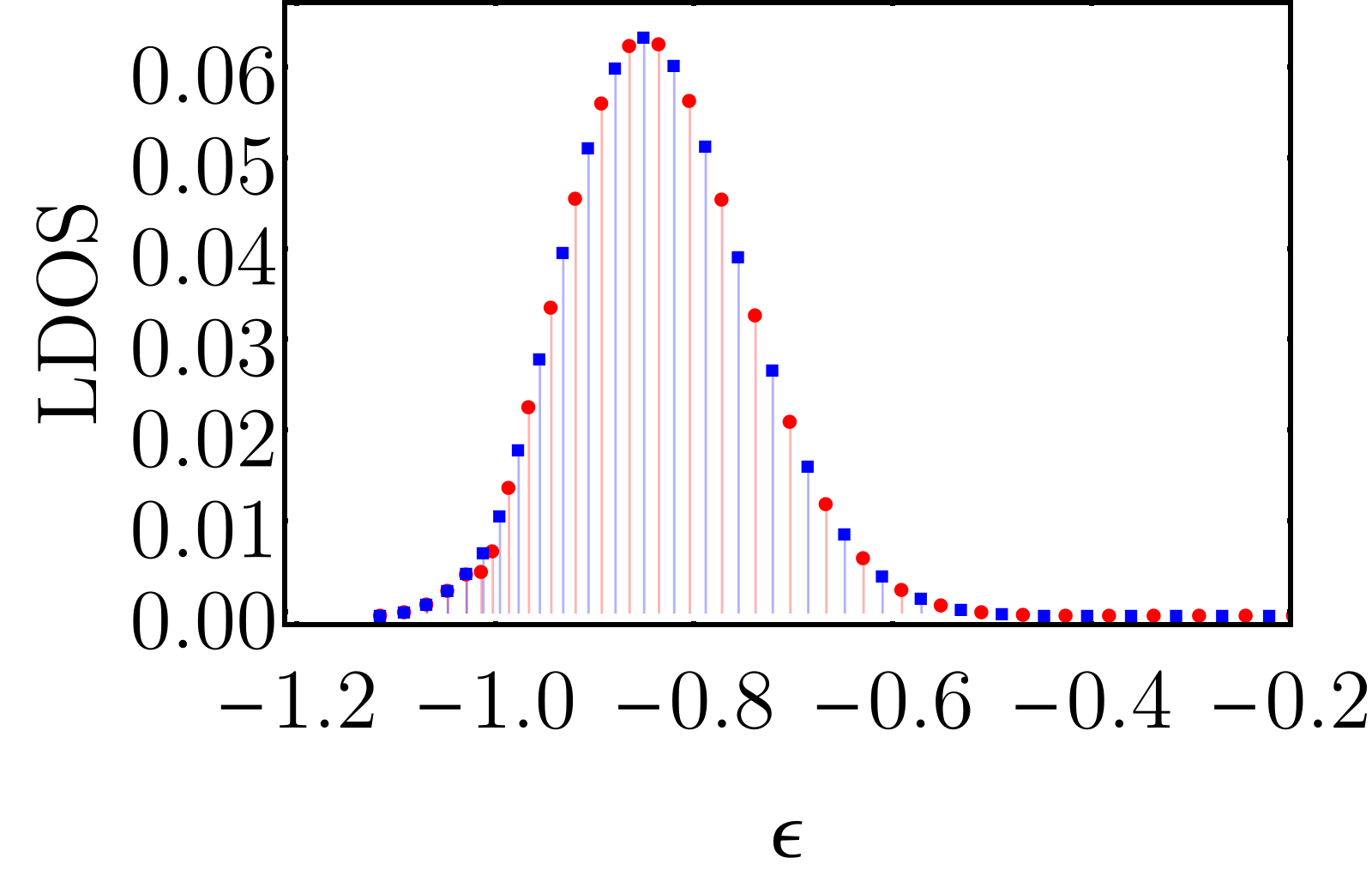} \\
    \end{tabular}
    \caption{Contour plot of $\log_{10}|\mathcal{Z}(\beta+it)|^{2}$ in the $(\beta,t)$ plane (top). Vertical red dashed lines signal the $t$-axis ($\beta=0$). The corresponding local density of states (bottom). Red circles correspond to the population coefficients of positive-parity states in the final Hamiltonian, $|c_{n,+}|^{2}$, while blue squares denote the same for negative-parity states, $|c_{n,-}|^{2}$. Three quenches $\lambda_{i}\to\lambda_{f}=1.6$ have been performed: under the ESQPT with with $\lambda_{i}=2.3$ (left column), onto the ESQPT with $\lambda_{i}=4$ (middle column), and above the ESQPT with $\lambda_{i}=20$ (right column). The parameters in the initial state~\eqref{eq:initialstate} are $\alpha=1/2$ and $\phi=0$. System size is $j=50$.}
    \label{fig:ZandLDOS}
\end{figure*}

In Fig.~\ref{fig:ZandLDOS} we focus on three quenches $\lambda_{i}\to\lambda_{f}=1.6$ where an initial state of the form Eq. \eqref{eq:initialstate} changes only due to the variation of $\lambda_{i}$, with $\alpha=0.5$ fixed. The shape and location of the LDOS is clearly different for $\lambda_{i}=2.3,4,20$: the longer the quench, the higher the average energy~\eqref{eq:finale}. For $\lambda_{i}=2.3$ as $\Delta\lambda=|\lambda_{i}-\lambda_{f}|$ is small, the after-quench average energy is $\langle\epsilon_{f}\rangle=-1.08<\epsilon_{c}$ according to~\eqref{eq:finale} and the quench only populates energy eigenstates close to the ground-state. For $\lambda_{i}=4$, $\langle\epsilon_{f}\rangle=-1=\epsilon_{c}$, eigenstates are populated on both sides of the ESQPT at $\epsilon_{c}$; as a consequence, $P_{+}(E)$ and $P_{-}(E)$ very approximately coincide only for eigenstates below $\epsilon_{c}$, and the ESQPT is signaled by a dip in the LDOS close to $\epsilon_{c}$~\cite{Santos2015,Santos2016,Corps2022,PerezFernandez2011}. Finally, for $\lambda_{i}=20$, $\langle\epsilon_{f}\rangle=-0.85$ and the quench mainly populates eigenstates above the ESQPT. The after-quench energy distribution has an impact on the location of the zeros of $\mathcal{Z}(\beta+it)$, and in particular it affects which chain of zeros intersects the $\beta=0$ line and at what times. As the final energy average increases, the overall pattern of zeros shifts towards higher $\beta$ values as visible in the contour plots of $|\mathcal{Z}(\beta+it)|^{2}$. The times when the survival probability rate function shows sharp peaks and whether or not these peaks occur in a quasiperiodic fashion is thus tightly connected with the LDOS.

\subsection{Relation to ESQPTs}

\begin{figure}[h!]
    \centering
    \begin{tabular}{c}
\hspace{0.3cm}\includegraphics[width=0.4\textwidth]{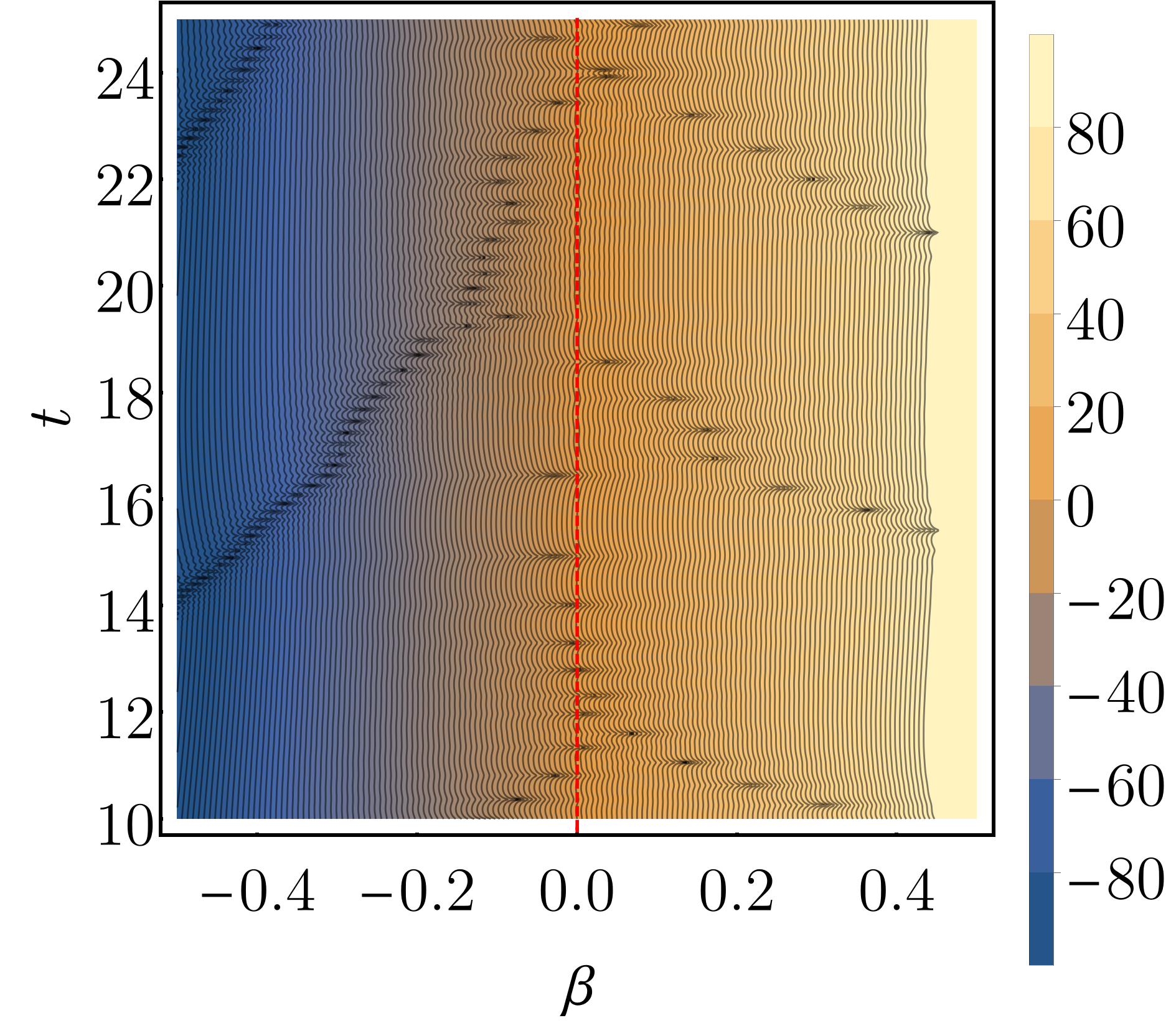} \\ \hspace{-1cm}\includegraphics[width=0.4\textwidth]{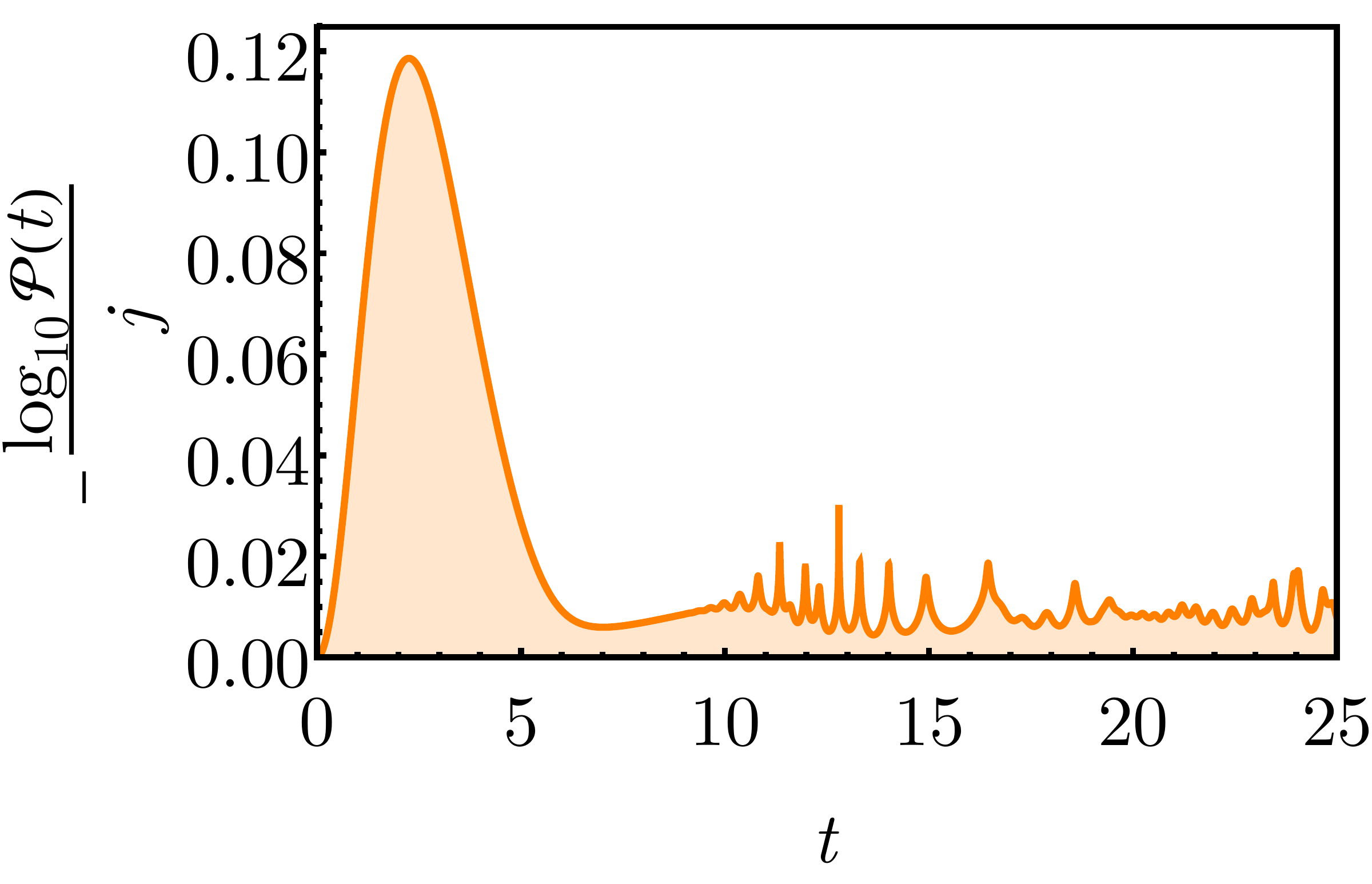} \\
    \end{tabular}
    \caption{Contour plot of $\log_{10}|\mathcal{Z}(\beta+it)|^{2}$ in the $(\beta,t)$ plane (top); system size is $j=200$. Vertical red dashed line signals the $t$-axis ($\beta=0$). The resulting rate function of the survival probability (bottom). The quench to the ESQPT $\lambda_{i}=4\to \lambda_{f}=1.6$ with $h=1$ has been performed. The parameters in the initial state~\eqref{eq:initialstate} are $\alpha=1/2$ and $\phi=0$. }
    \label{fig:ESQPT}
\end{figure}

ESQPTs manifest themselves as non-analyticities in the high-energy excitation spectrum of a many-body quantum system. This is a static effect with important dynamical consequences, some of which have been thoroughly studied~\cite{Kloc2018,Kloc2021,Santos2015,Santos2016,Wang2021,Corps2022PRA}. For systems with a single classical degree of freedom, the divergence of the density of states at the critical energy is transferred directly to the expectation values of certain observables in the energy eigenbasis~\cite{Puebla2016,Cejnar2021}. In the context of the survival probability, the ESQPT is expected to have some relevant impact, as can be inferred both from the quantum level flow diagram and, at the semiclassical level, from the phase space portrait (c.f. Fig.~\ref{fig:lipkinphase}). Indeed, it has been shown that quenching the initial state so that the final state significantly overlaps the ESQPT spectral region induces a variety of stabilizing or destabilizing effects In particular, the survival probability ceases to show the characteristic succession of revivals until it saturates to an asymptotic value~\cite{Relano2008,PerezFernandez2011,Kloc2018,Kloc2021}.

\begin{figure}[h!]
    \centering
    \begin{tabular}{c}
\hspace{0.3cm}\includegraphics[width=0.4\textwidth]{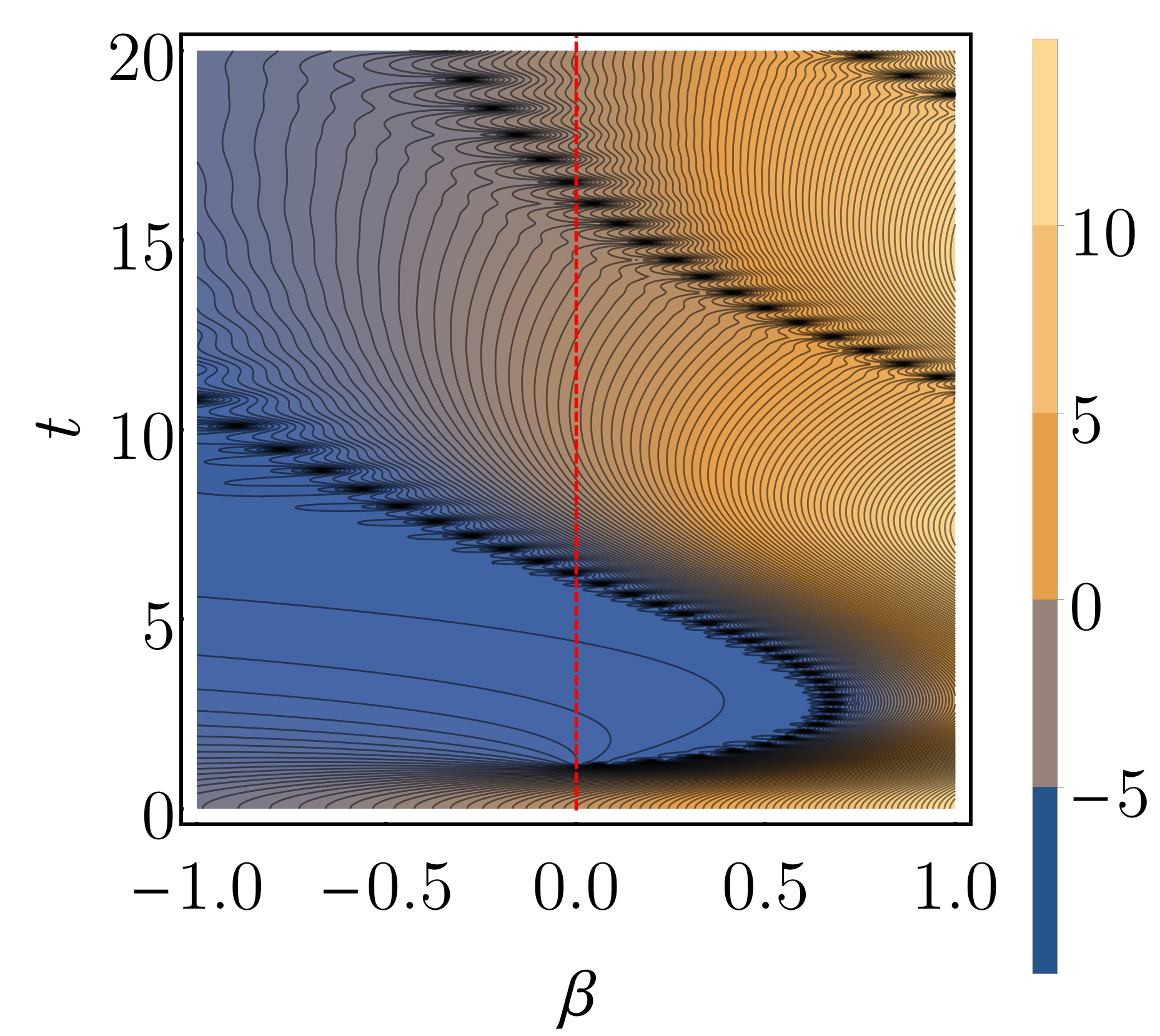} \\ \hspace{-1cm}\includegraphics[width=0.4\textwidth]{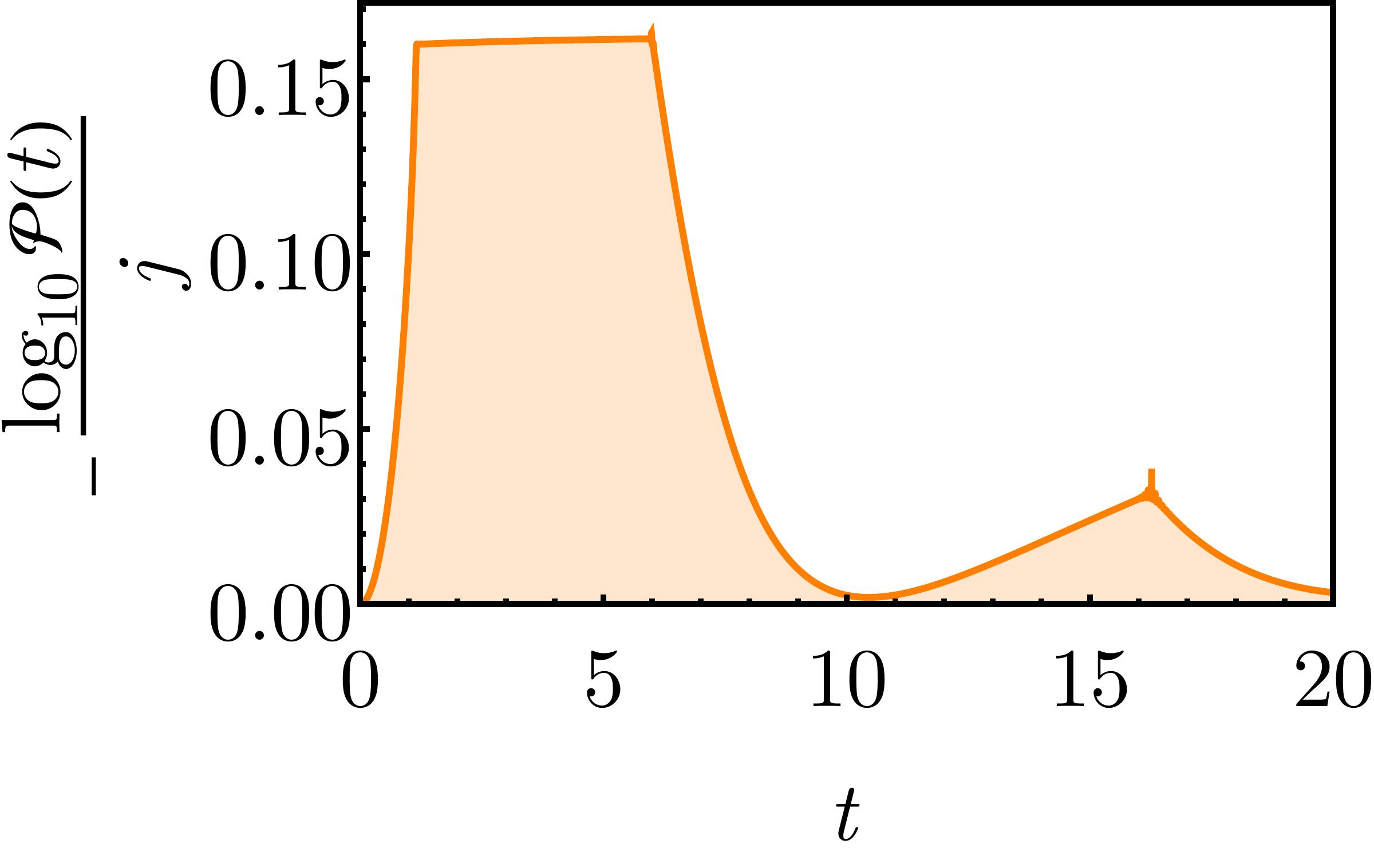} \\
    \end{tabular}
    \caption{Contour plot of $\log_{10}|\mathcal{Z}(\beta+it)|^{2}$ in the $(\beta,t)$ plane (top); system size is $j=50$. Vertical red dashed line signals the $t$-axis ($\beta=0$). The resulting rate function of the survival probability (bottom), for $j=400$. A quench $h_{i}=0.8\to h_{f}=0$ with $\lambda=1$ kept fixed has been performed. The parameters in the initial state~\eqref{eq:initialstate} are $\alpha=1/2$ and $\phi=0$. }
    \label{fig:relationesqpt}
\end{figure}

In Fig.~\ref{fig:ESQPT} we have drawn contour plots of ${|\mathcal{Z}(\beta+it)|^{2}}$ together with the corresponding rate function for the same quench presented in previous Fig.~\ref{fig:ZandLDOS}~(middle column), albeit for a larger system size. The quench leads the initial state to the ESQPT. The destabilized behavior of the rate function is clearly observed, and its overall pattern is qualitatively different from that obtained when quenching the initial state away from the ESQPT (compare, for example, with Fig.~\ref{fig:ZandSP}). From the viewpoint of the complex-time survival probability, we find that an extensive number of zeros cluster together around the $\beta=0$ line in a disorganized way. This behavior is again quite different from that in Fig.~\ref{fig:ZandSP}, where the zeros close to the $\beta=0$ line occur in a quasiperiodic way. It is noticeable that the zeros appearing on the far left and right of the $\beta=0$ line do show this seemingly organized pattern, so quenching the initial state sufficiently away from the ESQPT, as was described in the previous subsection, produces the standard behavior of the rate function previously observed in fully connected models~\cite{Corps2022,Corps2022arxiv, Homrighausen2017,Lang2018concurrence,Puebla2020}. 
The presence of ESQPTs is evidently imprinted in the distribution of the zeros in the complex-time survival probability.

It was shown in~\cite{Corps2022,Corps2022arxiv} that dynamical phase transitions of the first type, DPTs-I, can be directly attributed to the non-analytical change of shape of the classical version of the model and, therefore, to the ESQPT. The ESQPT thus separates two distinct dynamical phases characterized by some dynamical order parameters, which differ from the traditional order parameters of equilibrium phase transitions \cite{Sachdev1999}. With respect to the focus of the present work, DPTs-II, it was shown that there is a mathematical restriction on the non-analytical points of the parity-projected return probability, $\mathcal{L}(t)$, originally proposed in \cite{Heyl2014} for systems with symmetry-broken phases. This quantity is defined as $\mathcal{L}(t)=|\langle E_{0,+}(\Lambda_{i}^{\mu})|\Psi_{t}(\Lambda_{f}^{\mu})\rangle|^{2}+|\langle E_{0,-}(\Lambda_{i}^{\mu})|\Psi_{t}(\Lambda_{f}^{\mu})\rangle|^{2}\equiv \mathcal{L}_{+}(t)+\mathcal{L}_{-}(t)$, where it is assumed that the initial state is a certain superposition of the ground-state manifold of the pre-quench Hamiltonian, $\hat{H}(\Lambda_{i}^{\mu})$. Observe how this differs from the usual survival probability, which is just $\mathcal{P}(t)=|\langle \Psi_{0}(\Lambda_{i}^{\mu})|\Psi_{t}(\Lambda_{f}^{\mu})\rangle|^{2}$. For initial states of the form~\eqref{eq:initialstate}, one has~\cite{Corps2022,Corps2022arxiv}
\beq\label{eq:Lt}
\mathcal{L}(t)=\mathcal{L}_{+}(t)+\mathcal{L}_{-}(t)\equiv \alpha |f_{+}(t)|^{2}+(1-\alpha)|f_{-}(t)|^{2},
\eeq
where $f_{\pm}(t)=\sum_{n}|c_{n,\pm}|^{2}e^{-iE_{n,\pm}(\lambda_{f})t}$ and $c_{n,\pm}$ denote the expansion coefficients of the initial state in the final Hamiltonian eigenbasis. The survival probability in the same terms reads as
\beq\label{eq:Pt}
\mathcal{P}(t)=|\alpha f_{+}(t)+(1-\alpha)f_{-}(t)|^{2}.
\eeq
It is clear that $\mathcal{P}(t)$ keeps track of the interference between positive and negative parity sectors, while $\mathcal{L}(t)$ does not, as each of these contribute separately.

The non-analytical points in $\mathcal{L}(t)$ have been argued~\cite{Heyl2014} to result from crossing points of the functions $\Omega_{\pm}(t)$ that appear in $\mathcal{L}_{\pm}(t)=e^{-N\Omega_{\pm}(t)}$~\cite{Touchette2009}. It has been shown~\cite{Corps2022,Corps2022arxiv} that in the symmetry-broken phase (in our model, below the ESQPT line), eigenlevel degeneracies and the constancy of $\hat{\mathcal{C}}$ imply that $f_{+}(t)=f_{-}(t)$. This means that (i) \textit{no crossings} of the $\Omega_{\pm}(t)$ functions occur in the symmetry-broken phase and, therefore, this mechanism~\cite{Heyl2014} does not properly explain the appearance of kinks in its rate function in this spectral region; (ii) in the symmetry-broken phase, both return probabilities coincide in the infinite-size limit, $\mathcal{L}(t)=\mathcal{P}(t)$, while in the symmetry-restored phase (here, above the ESQPT critical line), these functions differ, $\mathcal{L}(t)\neq \mathcal{P}(t)$. In the symmetry-broken phase, the non-analytical kinks in $\mathcal{L}(t)$ are a consequence of the zeros of $\mathcal{P}(t)$, which, as we have seen (see e.g. Fig.~\ref{fig:ZandLDOS}), can very well occur near the $\beta=0$ line when the system is quenched to the symmetry-broken phase. These kinks, however, are completely unrelated to the mechanism of crossings in $\Omega_{\pm}(t)$, which can only happen in the symmetry-restored phase~\cite{Corps2022,Corps2022arxiv}. In this phase, the origin of non-analytical times in $\mathcal{L}(t)$ and $\mathcal{P}(t)$ is different, but the manifestations can be somewhat similar depending on the quench protocol chosen. 

On top of that, it has been proposed that the crossing of the ESQPT~\cite{Corps2022} is related to the appearance either regular or anomalous~\cite{Homrighausen2017} DPTs. A phase diagram has been proposed where in the symmetry-broken phase anomalous DPTs should be found, while in the symmetry-restored phase regular DPTs are obtained. By taking an extreme example, here we show that this is however not necessarily the case. In Fig.~\ref{fig:relationesqpt} we have represented the complex-time survival amplitude and the rate function of the survival probability for a quench not in $\lambda$, but in the magnetic field, ${h_{i}=0.8}\to{h_{f}=0}$ with $\lambda=1$ fixed. The average energy after the quench is $\langle\epsilon_{f}\rangle=-0.18<\epsilon_{c}$. The spectrum of the final Hamiltonian is completely degenerate in parity doublets because the ESQPT at $\epsilon_{c}=-h$ coincides with the upper boundary of the spectrum $\epsilon_{\mathrm{max}}=h$ at $h_{f}=0$. This is clearly shown in Fig.~\ref{fig:lipkinphase} (bottom) for a finite-size system, $j=50$. In Fig.~\ref{fig:relationesqpt}, three lines of zeros of $\mathcal{Z}(\beta+it)$ cross the $\beta=0$ line around $t\approx 1, 6, 16$, with the corresponding non-analytic points showing up in the rate function.  The first non-analyticity at $t\approx 1$ occurs even before the second revival of $\mathcal{P}(t)$, which means that this is a regular DPT, even though no ESQPT has ever been crossed in the process. The flatness of the rate function in the interval $1\lesssim t\lesssim 6$ can be understood from $\mathcal{Z}(z)$, which shows very weakly varying contour lines in this time interval. Note also that the first two non-analytic points in the rate function are of a different nature as those previously shown in this work. Instead of `peaks', the rate function displays `breaks'. Similar shapes have been previously observed~\cite{Lang2018concurrence}. 

The results of this section highlight that the ESQPT plays an important role in certain aspects of DPTs, for example establishing boundaries on the regions where certain mechanisms are allowed~\cite{Corps2022,Corps2022arxiv}, while other structural aspects of them, such as the classification in regular and anomalous DPTs, seem to escape this interpretation.

\section{Conclusions}\label{sec:conclusions}
Traditional thermal and ground-state quantum phase transitions can be understood via complexification of some relevant system parameters, such as the temperature or the value of an interaction strength. The ensuing non-analyticities, fully realized strictly in the infinite-size limit, show their traces, at finite size, in a complex parameter space. Based on this well-established notion, we have defined the complex-time survival amplitude, where the time variable is extended onto the complex domain. We have shown that dynamical phase transitions, understood as instants where an initial state is orthogonal to its time-evolved counterpart, are signaled through the zeros of the complex-time survival amplitude close to the real $t$-axis (imaginary $z$-axis). Such zeros are readily identified even for quite small system sizes, when precursors of criticality in the behavior of the survival probability itself can be ambiguous.

We illustrate these general ideas on the fully connected transverse-field Ising model excited in different ways far from its equilibrium state. We observe rich collections of zeros in the landscapes of the initial-state survival amplitude in the complex time domain. Transformations of these landscapes and the corresponding variations in the patterns of zeros allow us to easily identify and interpret qualitative changes in the system dynamics when certain parameters of the out-of-equilibrium protocol, such as details of the initial state or the average energy of excitation, are modified. The complex-time extension allows us to watch the origin and gradual development of dynamical critical structures in the evolution of the real survival amplitude.

We have also analyzed the influence of the excited-state quantum phase transition, which in the present and many other models separate a $\mathbb{Z}_{2}$ symmetry-broken phase and the corresponding symmetry-restored phase. Quenching a state initially prepared in the broken-symmetry phase to the excitation region below or above the critical energy line produces different patterns of zeros in the complex-time survival amplitude near the time axis and therefore leads to different sequences of dynamical phase transitions. Various mechanisms generating dynamical non-analyticities play role on both sides of the critical line. Quenching the broken-parity initial state onto the critical line produces a disorganized pattern of zeros near the time axis; as a result, the survival probability behaves erratically, most likely hiding the eventual signatures of dynamical phase transitions.

\begin{acknowledgments}
A. L. C. is grateful to Armando Rela\~{n}o and Rafael A. Molina for guidance and discussions. 
We also acknowledge discussions with Jakub Novotn\'{y}.  
This work has been supported by the Spanish grant PGC-2018-094180-B-I00 funded by Ministerio de Ciencia e Innovaci\'{o}n/Agencia Estatal de Investigaci\'{o}n MCIN/AEI/10.13039/501100011033 and FEDER ``A way of making Europe'', by the Grant No. 20-09998S funded by the Czech Science Foundation, and by the project UNCE/SCI/013 of the Charles University in Prague. 
A. L. C. acknowledges support from `la Caixa' Foundation (ID 100010434) through the fellowship LCF/BQ/DR21/11880024, which financed his scientific stay in Charles University in Prague leading to this work.
\end{acknowledgments}

\end{document}